\documentclass[preprint,12pt,authoryear]{elsarticle}

\usepackage{amssymb}
\usepackage{natbib}
\usepackage{lineno,hyperref}
\usepackage{amsmath}
\usepackage{comment}
\usepackage{tablefootnote}

\modulolinenumbers[2]

\journal{New Astronomy}
\begin{document}

\begin{frontmatter}



\title{New models of reflection spectra for terrestrial exoplanets: Present and prebiotic Earth orbiting around stars of different spectral types}


\author[inst1,inst2]{Manika Singla}

\cortext[mycorrespondingauthor]{Corresponding author}
\ead{manika.singla@iiap.res.in}

\author[inst1]{Sujan Sengupta}

\affiliation[inst1]{Indian Institute of Astrophysics, Koramangala, Bengaluru, 560034, India}
            %addressline={II Block, Koramangala}, 
            %city={Bengaluru},
            %postcode={560034}, 
            %state={Karnataka},
            %country={India}
     
\affiliation[inst2]{Department of Physics, Pondicherry University, Kalapet, Puducherry, 605014, India}
            %addressline={ R.V. Nagar, Kalapet}, 
            %city={Puducherry},
            %postcode={605014}, 
            %country={India}       

% These dates will be filled out by the publisher
%\date{Accepted XXX. Received YYY; in original form ZZZ}

% Enter the current year, for the copyright statements etc.
%\pubyear{2022}

%\DeclareUnicodeCharacter{2212}{-}
%\begin{document}
%\label{firstpage}
%\maketitle

%%%%%%%%%%%%%%%%%%%%%%-----------------abstract-------------------%%%%%%%%%%%%%%%%%%%%%%%%%%

\begin{abstract}
In order to recognize a habitable exoplanet {from future observed spectra}, we present new model reflected spectra and geometric albedo for modern and prebiotic ($\sim$3.9 Ga) Earth-like exoplanets orbiting within the habitable zone of stars of spectral types F, G, K and M. We compute this for various atmospheric and surface compositions of the planets. Molecules that are potential biosignatures and act as greenhouse agents are incorporated in our model atmosphere. Various combinations of solid and liquid materials such as ocean, coast, land consisting of trees, grass, {sand or rocks determine the surface} albedo of the planet. Geometric albedo and model reflected spectra for {a set of nine potential habitable planets, including Proxima Centauri b, {TRAPPIST}-1d, Kepler-1649c and Teegarden's Star-b, are also} presented.
We employ the opacity data derived by using the open-source package {\tt Exo-Transmit} and adopt different {atmospheric} {Temperature-Pressure} profiles depending on the properties of the terrestrial exoplanets. The {model-reflected} spectra are constructed by numerically solving the multiple scattering radiative transfer equations. We verified our model reflected spectra for a few specific cases by comparing with {those} published by other researchers. We demonstrate that prebiotic Earth-like exoplanets and present Earth-like exoplanets with increased amount of greenhouse gases in their atmospheres scatter more starlight in the optical.
We also present the transmission spectra for modern and prebiotic Earth-like exoplanets with cloudy and cloudless atmospheres.
\end{abstract}

%\includnegraphics{grabs}


\end{frontmatter}

%%%%%%%%%%%%-------------------introduction----------------------%%%%%%%%%%%%%%%%%%%%%%%%%

%\linenumbers
\section{Introduction} \label{sec:intro}
Since the discovery of 51 Pegasi b \citep{mayor1995jupiter}, more than {5000 exoplanets candidates have been discovered using} various methods, {yet a little has been investigated about their atmospheres.} According to \cite{bryson2020occurrence}, around half of the Solar-type stars in our Galaxy might host rocky and potentially habitable planets within their habitable zones. 
But still we are far from finding any exoplanet that may {have an ambient} environment similar to that of the Earth. We will be a step closer to finding out such planets if we can characterize the atmospheres {of terrestrial exoplanets} \citep{2004ASPC..321..170S, morley2015thermal, kaltenegger2017characterize, Alonso2018, kopparapu2020characterizing, quanz2021atmospheric}.

The classical circumstellar habitable zone is defined as the region around a star where the surface temperature of a planet is appropriate for water to exist in liquid state \citep{huang1959occurrence, huang1960life, whitmire1991habitable, kasting1993habitable, kopparapu2013habitable}. A few of the planets {discovered by the NASA's {\it Kepler} space mission,} possibly located in the habitable zone of their host stars {\citep{covone2021efficiency}, are} also of great interest. {The recent TESS (Transiting Exoplanet Survey Satellite) discoveries include Super-Earth and Sub-Neptunes orbiting around HD 108236 (G3V), GJ 3929 b, which is a hot Earth sized planet orbiting around M3.5V star \citep{daylan2021tess, kemmer2022discovery}. Many Earth-like planets, including TOI-700d, which lie in their host star's habitable zone, discovered by TESS are also important \citep{kaltenegger2021around}. Many potentially habitable exoplanets have also been discovered by RV Spectrographs \citep[and many more]{jurgenson2016expres, wildi2017nirps}. }

When stellar radiation is incident on the surface of a planet, parts of it get reflected, absorbed and transmitted depending on the wavelength of the radiation and the angle of incidence of the stellar flux {\citep{seager2010exoplanet, perryman2018exoplanet}}. The planetary reflected {spectra are generated} by the fraction of the incident stellar radiation reflected along our line of sight  \citep{selsis2008terrestrial}. The interaction of the incident stellar radiation with the matter in the upper atmosphere of the planet introduces signatures of the atmospheric chemical composition in the reflected spectra. {However, when an} exoplanet transits across the host star, a tiny portion of the stellar disk is blocked yielding into a reduction in the stellar flux. At the same time, a fraction of star-light  passes through the planetary atmosphere, if any, and brings the information on the chemical composition of the planetary atmosphere. This is known as the {transmission spectrum} \citep{palle2009earth, wunderlich2019detectability}.  
The signatures of the molecules present in the planetary atmosphere are revealed in the absorption features  of reflected and transmitted spectra \citep{Tinetti}. 
 {If a combination of biosignatures, such as oxygen, ozone, water and methane, were detected} in the atmosphere of rocky exoplanets in habitable zone, {there would be a high possibility} that the planet harbours life \citep{owen1980search, sagan1993search, selsis2004atmosphericbiomarkers, scharf2009extrasolar, grenfell2014sensitivity, fujii2018exoplanet, claudi2019biosignatures}.

 {Previous studies suggested} that the potentially habitable planets can orbit {stars of F, G, K and M spectral types} \citep{selsis2000physics}. According to \cite{kasting1993habitable}, the most potentially habitable planets orbit around late F, G and early K-type main-sequence stars. Stars whose spectral type is earlier than F0 have less than 2 Gyr main sequence lifetimes and {hence are very less} likely to have planets that can harbour life \citep{segura2003ozone}. On the other hand, {M dwarfs} have much less probability of having life supporting planets orbiting around them because their habitable zones are much nearer and narrower and so the planets in the habitable zone are exposed to strong UV radiation and strong flares \citep{huang1959occurrence,huang1960life, hart1979habitable}. {Also most planets in the inner habitable zone of M dwarfs are tidally locked  \citep{kasting1993habitable, segura2003ozone, martinez2019exomoons}} and the permanent day or night {side of the planet may have} hostile environment for life. 
 {Nearly 70$\%$ of all stars in our Galaxy are M dwarfs and rocky} planets orbiting {M dwarfs} may be the most common in the universe {{\citep{henry2006solar, shields2016habitability, meadows2018habitability, lin2020high, reyle202110, sabotta2021carmenes}}.} Therefore, it is important to include the exoplanets orbiting {M dwarfs} as well in any investigation and probe. 

Exoplanets similar to the prebiotic Earth {($\sim$3.9\,Ga)} can also be the potential candidates for supporting life on them. Prebiotic Earth contained no {free molecular oxygen but} carbon dioxide and nitrogen as the most dominant gases in their atmospheres \citep{rugheimer2015uv}. Prevalent oxygenation of the Earth's atmosphere took place somewhere between {2.45\,Ga and 2.32\,Ga, which} is known as the Great Oxidation Event {GOE} \citep{holland2002volcanic, bekker2004dating, guo2009reconstructing}. Discovery of the {biomarkers} in sedimentary rocks {(banded iron formation)} with age {2.7\,Ga to 2.8\,Ga, which} are characteristic {of photosynthetic cyanobacteria,} indicates the appearance of O$_2$ in the Earth's atmosphere \citep{brocks1999archean}. Before this period, life survived through anoxygenic photosynthesis process. {The second oxygenation event took place around {0.8\,Ga to 0.5\,Ga}, which is known as Neoproterozoic Oxygenation Event (NOE). During that period, oxygen probably accumulated to the levels that are required for the animal life \citep{shields2011case, och2012neoproterozoic, hiatt2020iron}.}

 {About three decades ago, {the {\it Galileo} space mission obtained} the reflected spectra of the Earth over a relatively clear sky region of the Pacific Ocean, {north of Borneo, which was analysed by \cite{sagan1993search}}.
 {Previously, many groups have characterized the atmospheres of modern and prebiotic Earth-like exoplanets by calculating reflection and transmission spectra \citep{ehrenreich2006transmission, kaltenegger2009transits, kitzmann2010influence, domagal2014abiotic, wunderlich2019detectability, kaltenegger2020finding, lin2021differentiating}.
Studies have also been done for the {Earth-like} planets orbiting F, G, K and M stars \citep{segura2005biosignatures, grenfell2007response, rugheimer2013spectral, rugheimer2018spectra}.}
An open source radiative transfer model {\tt PICASO} to calculate the reflected spectra of exoplanets was presented by \cite{batalha2019exoplanet}.
Earth's transmission spectra through lunar eclipse {observations were calculated by \cite{palle2009earth, palle2010observations} and \cite{yan2015high}.}}

 {Previously, \cite{kreidberg2016prospects}, \cite{article}, \cite{dong2017proxima, lovis2017atmospheric, luger2017pale, meadows2018habitability, lin2020high, scheucher2020proxima} have characterized the atmosphere for Proxima Centauri b and \cite{de2018atmospheric, krissansen2018detectability, moran2018limits, zhang2018near, lustig2019detectability, hori2020trappist, lin2020high, turbet2020review, wunderlich2020distinguishing, may2021water} have extensively discussed about {TRAPPIST-1} system and in particular the planets {TRAPPIST}-1d and e. \cite{kaltenegger2013water}, on the other {hand, modeled} the transmission spectra for the planet Kepler-62e.
Clouds also play a crucial role in determining {reflection and transmission} spectra \citep{kitzmann2010clouds, kitzmann2010influence, kitzmann2011clouds, kitzmann2011clouds3, kawashima2019theoretical}.
 {\cite{fauchez2019impact} demonstrated} the effect of clouds and hazes on the transmission spectra of the planets in the habitable zone of {TRAPPIST-1, and} \cite{pidhorodetska2020detectability} worked on detectability of molecules through transmission spectroscopy.}

In this paper, we present the new synthetic reflected spectra of exoplanets similar to the modern and prebiotic Earth orbiting around stars of F, G, K and M spectral types. {If} the atmosphere is optically thick at pressure level much smaller than {10$^3$\,mbar}, most of the incident stellar radiation in the optical wavelength region {will get} absorbed and reflected only by the planetary atmosphere. However, the reflecting properties of the surface play a crucial role in the re-emission of thermal radiation at the {infrared} wavelength region. The surface albedo of solid {or liquid (ocean) surface} is also {considered, which} provides better and realistic model spectra. We also calculate the spectra for nine Earth-like planets that lie in the habitable zone of their host stars.

We also present model transmission {spectra for simulated terrestrial exoplanets} with atmospheric composition similar to that of the modern as well as prebiotic Earth. For this purpose, we {use the publicly} available software package {\tt Exo-Transmit}\footnote{\url{https://github.com/elizakempton/Exo_Transmit}} \citep{kempton2017exo}. A comparative study of the transmission spectra {calculated using the {\tt Exo-Transmit} code and the {\tt TauREx} software package \citep{waldmann2015tau} was presented} in \cite{sengupta2020optical}.

In the next {Section}, we discuss the methodologies adopted to calculate the reflected spectra and the validation of our results. In particular, numerical methodologies are discussed in {Section} \ref{subsec:num} and the absorption and scattering opacities that are employed are described in {Section} \ref{subsec:opac} and we discuss about Temperature-Pressure profile in Section \ref{subsec:T-P} . Results are presented in {Section} \ref{sec:analysis_and_res}. The model reflected spectra {for both cases} - modern and prebiotic Earth-like exoplanets orbiting around stars of F, G, K and M spectral types are presented in {Section} \ref{sec:reflected_spectra} and the model reflected spectra of specific and interesting habitable terrestrial {planets are shown in} {Section} \ref{sec:case_studies}. The model transmission spectra are presented in {Section}  \ref{sec:trans_spectra}. Finally we discuss our results with specific conclusions in {Section} \ref{sec:disc_conc}.

%%%%%%%%%%%%%%%%5-----------------numerical methodology-----------%%%%%%%%%%%%%%%%%%%%%%%%%

\section{Methodology}\label{sec:section2}
\subsection{Numerical Methodology}\label{subsec:num}
To calculate the reflected spectra, we solved the multiple-scattering radiative transfer {equation} for diffused reflection and {transmission, which} for a plane-parallel geometry and azimuthal symmetry, is given by \citep {chandrasekhar1960radiative, sengupta2020optical}:
     
\begin{equation}\label{multscateqn}
\begin{split}
\mu\frac{dI(\tau,\mu,\lambda)}{d\tau}=I(\tau,\mu,\lambda)-
\frac{\omega}{2}\int_{-1}^1{p(\mu,\mu')I(\tau,\mu',\lambda)\text{d}\mu'}\\
-\frac{\omega}{4}F(\lambda) e^{-\tau(\lambda)/\mu_0}p(\mu,\mu_0)
\end{split}
\end{equation}

\noindent {where} $I(\tau,\mu,\lambda)$ is the specific intensity of the diffused radiation field  along the direction $\mu=\cos\theta$, $\theta$ being the angle between the axis of symmetry and the ray path,  $\omega$ is the albedo for single scattering, { $F(\lambda)$} is the incident stellar flux in the direction $-\mu_0$ and $\tau$ is the optical depth such that $d\tau=-\chi dz$, where $\chi$ is the total absorption {coefficient} or extinction coefficient, i.e., the sum of true absorption and absorption due to scattering.
In the above equation, $p(\mu,\mu')$ is the scattering phase function that describes the angular distribution of the
 {photons} before and after scattering. The scattering phase function depends on the nature of the scatterers. For scattering by atoms and molecules, the angular distribution is described by Rayleigh scattering phase function and is given by \cite{chandrasekhar1960radiative},
\begin{eqnarray}
p(\mu,\mu')=\frac{3}{4}[1+\mu^2\mu'^2+\frac{1}{2}(1-\mu^2)(1-\mu'^2)],
\end{eqnarray}   
where $\mu$ and $\mu'$ are the cosine of the angle before and after scattering with respect to the normal. 

As {pointed out by} \cite{sengupta2020optical}, in a scattering medium, the radiation field has two components: the {reflected and transmitted intensities, which} suffer one or more scattering {process,} and the directly transmitted {flux, which} is known as the reduced incident flux \citep{chandrasekhar1960radiative}, $\pi F(\lambda)e^{-\tau(\lambda)/\mu_0}$ in the direction $-\mu_0$. So, the reflected and the transmitted intensities {that are incorporated} through the second term in the right hand side of the above equation, {do not} include the reduced incident {flux, which} is described by the third term. 

We solved the multiple scattering radiative transfer equations by using the discrete space {theory that was} developed by \cite{peraiah1973numerical} and \cite{peraiah2002introduction}. The numerical code has extensively been used to solve the vector radiative transfer equations to incorporate scattering polarized spectra of brown dwarfs and self-luminous exoplanets {\citep{sengupta2009multiple, sengupta2010observed, sengupta2016detecting, marley2011probing, sengupta2016polarimetric,  sengupta2018polarization}}.  For the present work, we used the scalar version of the same numerical code by using the following steps:
\begin{enumerate}
\item As the vertical atmosphere is heterogeneous with respect to temperature, pressure and optical depth, we {divided} it into many {``shells"} of small optical depths. The thickness of each shell is less than or equal to a critical thickness $\rm{\tau_{critical}}$, which is calculated on the basis of the physical characteristics of the medium. If $\tau \leq \rm{\tau_{critical}}$, the reflection and the transmission {operators have} non-negative elements \citep{peraiah2001introduction}. We assume a constant temperature and pressure over each {shell, and} then integrate the radiation over all the shells.
\item The integration of the transfer equation is performed {on the shells,} which is a {two-dimensional} grid bounded by [r$_n$, r$_{n+1}$]$\times$[$\mu$$_{j-1/2}$, $\mu$$_{j+1/2}$], where, r$_n$ is the radial grid and $\mu_{j+1/2}$ is the angular {grid:}
\begin{eqnarray}
    \mu_{j+1/2} = \sum_{k=1}^{j} c_k , j = 1,2,...,\rm{J}
\end{eqnarray}
Here, c$_k$ are the weights of {Gauss-Legendre} quadrature formula. {We used} the plane-parallel approximation by making the shell curvature equal to zero.
\item The Gauss' quadrature formula is given as \citep{chandrasekhar1960radiative}:
\begin{equation}
    \int_{-1}^{+1} f(\mu) d\mu = \sum_{j=1}^m a_j f(\mu_j)
\end{equation}
where $\mu_1$,....., $\mu_m$ are the zeros of P$_m(\mu)$ and
\begin{equation}
    a_j = \frac{1}{P_{m}^{'}(\mu_j)}\int_{-1}^{+1} \frac{P_m(\mu)}{\mu-\mu_j} d\mu
\end{equation}
where, P$_m(\mu)$ is known as the Legendre Polynomial of order $m$. We used the 8-point Gauss' Quadrature Formula.
\item We obtained the transmission and reflection operators of the shell by comparing these discrete equations with the canonical equations of the interaction principle, which relates the {incident and emergent} radiation from a medium of given optical depth.
\item Combining all the shells by star algorithm { \citep{peraiah2002introduction}}, we obtained the total radiation field. Star algorithm combines the radiation for two consecutive shells {by putting them together and calculating the radiation field as a whole}.
\end{enumerate}
The numerical method has been described in detail in \cite{peraiah1973numerical}, \cite{peraiah2002introduction}, \cite{sengupta2009multiple} and \cite{sengupta2020optical}. 

\begin{figure}
\centering
\includegraphics[width=\linewidth]{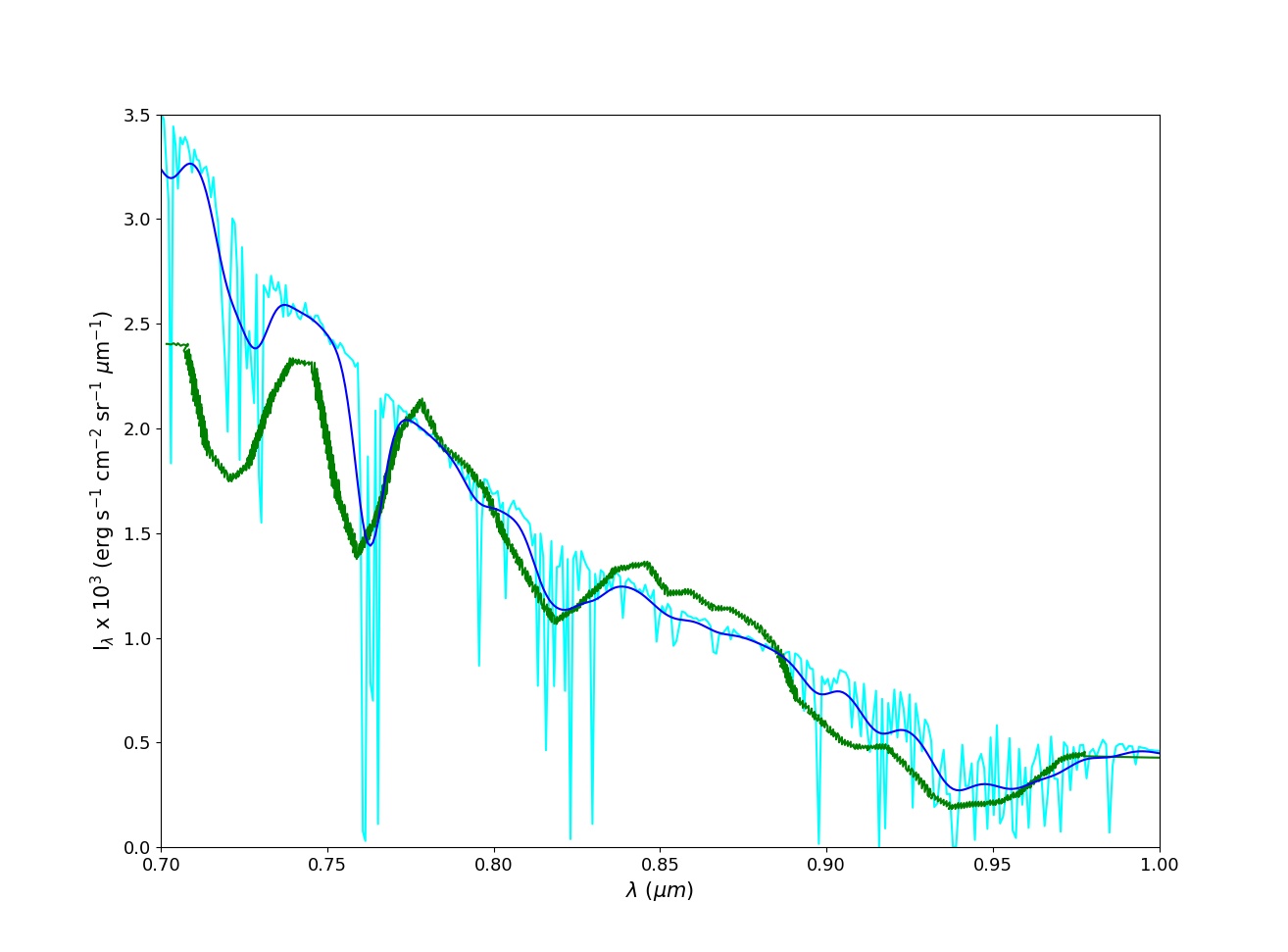}
\caption{Comparison of the model reflected spectrum (cyan) for an Earth-like exoplanet orbiting {around a Sun-like} star with the observed reflected spectrum (green) for the Earth obtained by {\tt {Galileo}} spacecraft \citep{sagan1993search}. {Blue color represents the model spectra at a spectral resolution same as NIMS in Galileo spacecraft.}}
\label{fig:1}
\end{figure}

In order to validate our numerical calculations, we compared the model reflected spectra of a terrestrial exoplanet orbiting around a {solar-type star} with the observed reflected spectra of the Earth \citep{sagan1993search} obtained by the {\it {Galileo}} {spacecraft}. This is presented in {Figure} \ref{fig:1}. We found an overall good match of the observed {low-resolution} spectrum with our model spectrum, in particular the dominant {water and oxygen bands}. The intensity decreases with wavelength in the infrared {region because} of the nature of input solar spectra and Rayleigh scattering (which is proportional to $\lambda^{-4}$).

\begin{figure}
\centering
\includegraphics[width=\linewidth]{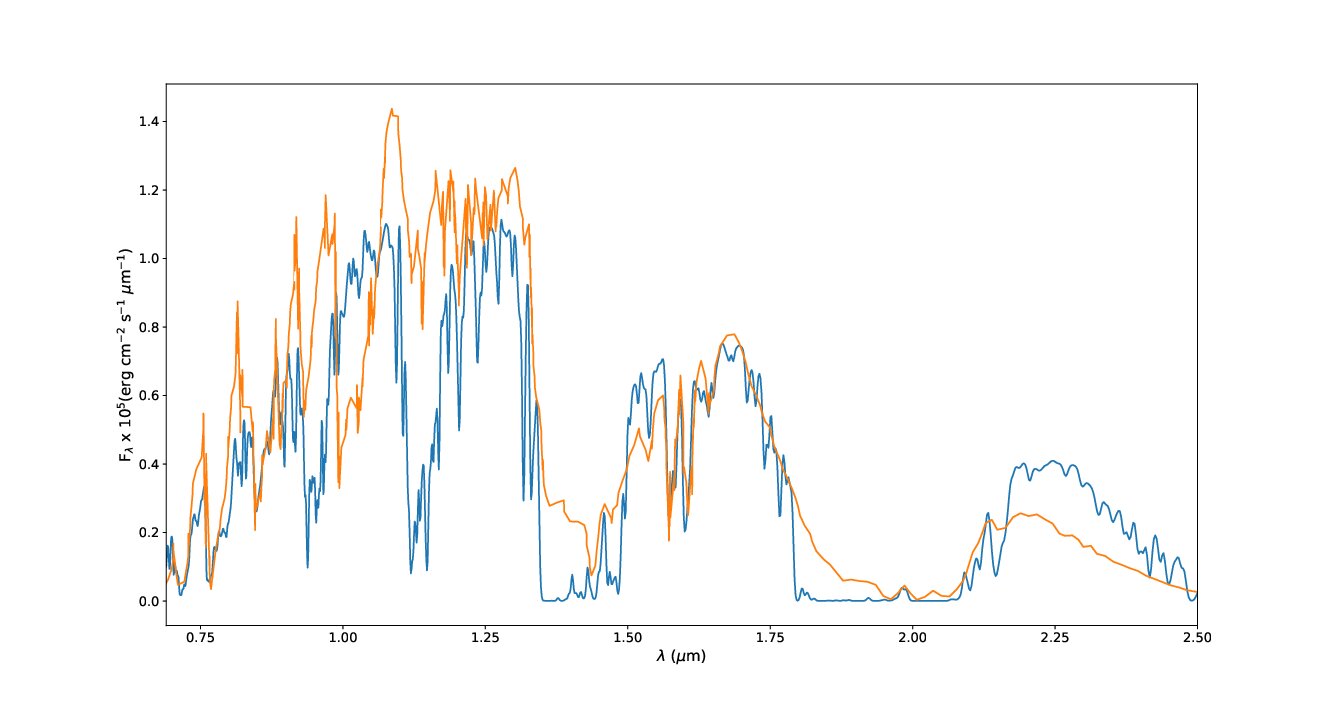}
\caption{Comparison of the modeled reflected spectra for {TRAPPIST}-1e (blue) with the spectra calculated by \citep{lin2020high} (orange).}
\label{fig:trap_veri}
\end{figure}

\begin{figure}
    \centering
    \includegraphics[scale=0.3]{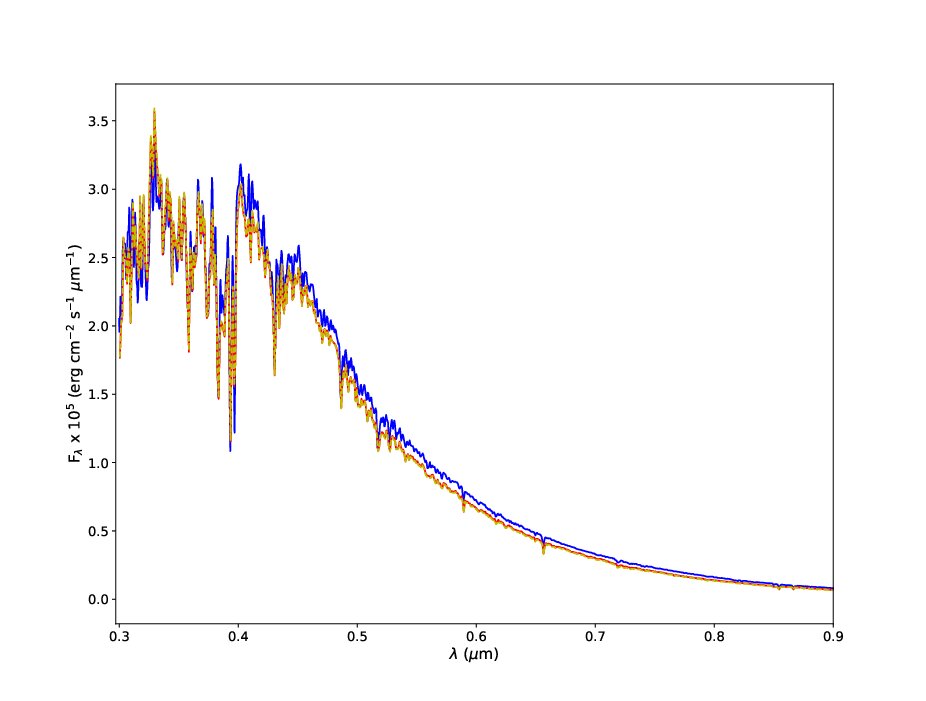}
    \includegraphics[scale=0.3]{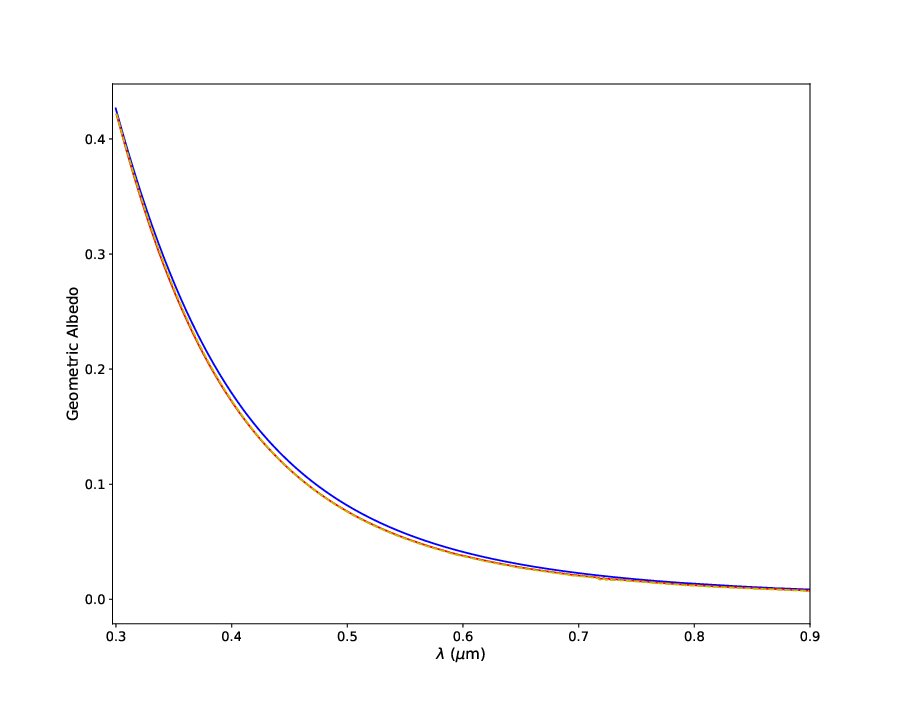}
    \caption{(a) Comparison of our model reflected spectrum (blue) calculated by using opacity from {\tt Exo-Transmit} package and observed atmospheric $\rm{ {T}}$-$\rm{ {P}}$ profile of the Earth with the theoretical spectrum (dashed yellow) provided by  S. Ranjan (priv. comm.) as well as with the spectrum calculated by our radiative transfer code using the same opacity and $\rm{ {T}}$-$\rm{ {P}}$ profile as used by S. Ranjan (red) for a prebiotic Earth-like exoplanet orbiting around a solar-type of star. {Yellow and red curves are essentially the same.} (b) Comparison of the geometric albedo for the above three cases.}
    \label{fig:2}
\end{figure}

 {Figure} \ref{fig:trap_veri} shows the comparison between the reflected spectra for {TRAPPIST}-1e as calculated by our model with that calculated by {\cite{lin2020high}}. We used the {same chemical abundance as in} \cite{lin2020high} {for verification purposes}. Temperature-Pressure ($\rm{T}$-$\rm{P}$) profiles employed for this {case were the} same as considered by \cite{o2019lessons}. Here also, the overall nature {is the same} and the slight variations are due to different opacities used {and also they used vertically variable atmospheric abundance, but we used vertically homogenous atmospheric abundance}.
We also compared our model reflected spectrum for prebiotic Earth-like exoplanets orbiting {around Sun-like stars} (atmosphere composed of only N$_2$ and {CO$_2$; \citealt{rugheimer2015uv})} with the model {spectrum} calculated by S. Ranjan {(priv. comm.). Figure \ref{fig:2}a} demonstrates that the reflection spectrum of prebiotic terrestrial exoplanets calculated by us matches very well with that derived by S. Ranjan. The slight variation is again due to the differences in opacities used {in both models.  Figure} \ref{fig:2}b shows the comparison of the geometric {albedo, which} also matches very well. Here there are no absorption lines because the {considered molecules} show absorption beyond the limit of the wavelength considered {in this work (2.49\,$\mu$m)}.

%%%%%%%%%%%%%%%%-------------absorption and scattering opacity----------%%%%%%%%%%%%%%%%

\subsection{Absorption and Scattering {Opacity}}\label{subsec:opac}

In order to calculate the reflection and the transmission spectra,  we {calculated} the absorption and scattering coefficients of the atmosphere by using the {\tt Exo-Transmit} software package \citep{kempton2017exo}. In this package, the opacities for 30 molecular and atomic species on a fixed temperature-pressure-wavelength grid are tabulated. The wavelength grid ranges from 0.3 to {30\,$\mu$m} {at low spectral resolution of $\mathcal{R} \approx$ 1000}.
The temperature and pressure range at which the absorption and scattering coefficients {were calculated} for each wavelength were {100--3000\,K and $10^{-6}$--$10^{6}$\,mbar} respectively. The opacities {were} derived by using the line lists given {by \cite{lupu2014atmospheres}}. The gas opacities {were} adopted from the widely used database of \cite{ freedman2008line, freedman2014gaseous}.  The individual opacity sources are the atomic and molecular opacity weighted by their abundances and the total Rayleigh scattering opacity. Since the Earth's atmosphere is sufficiently cool, {we neglected} the collision induced absorption of hydrogen. {We adopted} the molecular abundances of the present Earth \citep{sagan1993search} as described in Table \ref{table}. 
\begin{table}
\centering
\begin{tabular}{ |c|c|c|c|} 
\hline
Molecule & Abundance (volume mixing ratio) \\
\hline
N$_2$ & 0.78 \\ 
O$_2 $ & 0.21\\
H$_2$O & 0.03 - 0.001 \\ 
Ar & 9$\times$10$^{-3}$ \\
CO$_2$ & 3.5$\times$10$^{-4}$ \\
CH$_4$ & 1.6$\times$10$^{-6}$ \\
N$_2$O & 3$\times$10$^{-7}$ \\
O$_3$ & 10$^{-7}$--10$^{-8}$\\
\hline
\end{tabular}
\caption{Molecular abundance for the Earth's atmosphere \citep{sagan1993search}. }
\label{table}
\end{table} 
For the prebiotic Earth-like exoplanets, {we considered} a carbon dioxide dominated atmosphere with the molecular abundance as 10$\%$ CO$_2$, trace {amount}s of CH$_4$ and the remaining N$_2$ as considered {by} \cite{kaltenegger2007spectral}.
The atmospheric abundances of Proxima Centauri b, Kepler-442b, Kepler-62e, Kepler-22b, Kepler-1649c, TOI-400d, Teegarden's {Star} b, Trappist-1d and {TRAPPIST}-1e {were} considered to be the same as that of the present {Earth. For} the verification purpose for {TRAPPIST}-1e, the atmospheric abundance and $\rm{ {T}}$-$\rm{ {P}}$ profile {were} adopted following \cite{o2019lessons}.
The molecular abundances {were} incorporated in the equation of states file in the {\tt Exo-Transmit} {package, in which} pressure and temperature {were} also included for the atmospheric layers.

\subsection{Temperature-Pressure profile}\label{subsec:T-P}
For the terrestrial exoplanets, the internal temperature is negligible as compared to the irradiated temperature. Thus the incident stellar flux at the top-most layer of the atmosphere and the molecules present in the atmosphere determine the Temperature-Pressure ($T$-$P$) profile of the terrestrial exoplanets.

\begin{figure}
\centering
\includegraphics[width=\linewidth]{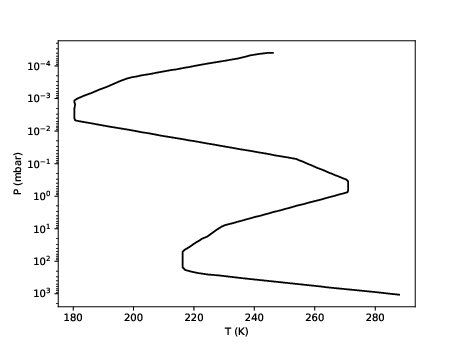}
\caption{Atmospheric Temperature-Pressure (T-P) profile for the present Earth \citep{atmosphere1976us}.}
\label{fig:tp}
\end{figure}

Most of the stellar radiation gets reflected from the upper layers of the planetary atmosphere. So, the temperature structure of the outer layers of the atmosphere mostly determines the reflected spectra. In the case of transmission spectra also, the lower atmosphere cannot be probed for most of the wavelengths \citep{kaltenegger2009transits}. Hence, for the present Earth-like exoplanets orbiting around {stars of} F, G, K and M spectral types, we adopted the Earth's atmospheric $T$-$P$ profile \citep{atmosphere1976us} {, which} is shown in {Figure} \ref{fig:tp}. As we go upwards from the solid surface of the Earth, the temperature decreases continuously with the decrease in pressure, thus following the ideal gas law in the tropospheric region. This region extends roughly about {9 km at the poles and 17 km at the equator \citep{caballero2022detailed}. We will roughly consider the height of the tropopause equal to 11 km for our calculations.} In the stratospheric region {, which} extends about 35 km above the tropopause, the temperature increases with the decrease in pressure due to the presence of ozone gas {, which} absorbs the ultraviolet radiation. This is known as the temperature inversion \citep{atmosphere1976us}.
The $T$-$P$ profile for the atmospheres of early Earth-like exoplanets is taken the same as considered in \cite{kaltenegger2007spectral} for Epoch 0 (3.9 Ga).

\section{Results and Analysis}\label{sec:analysis_and_res}

\subsection{The Reflection Spectra}\label{sec:reflected_spectra}
While orbiting its host star, an exoplanet reflects part of the starlight along our line of sight. We observe the maximum reflected radiation when the planet is almost at full phase or at zero degree phase angle, i.e,  just before or after the secondary eclipse position.

\begin{figure}
    \includegraphics[width=\linewidth]{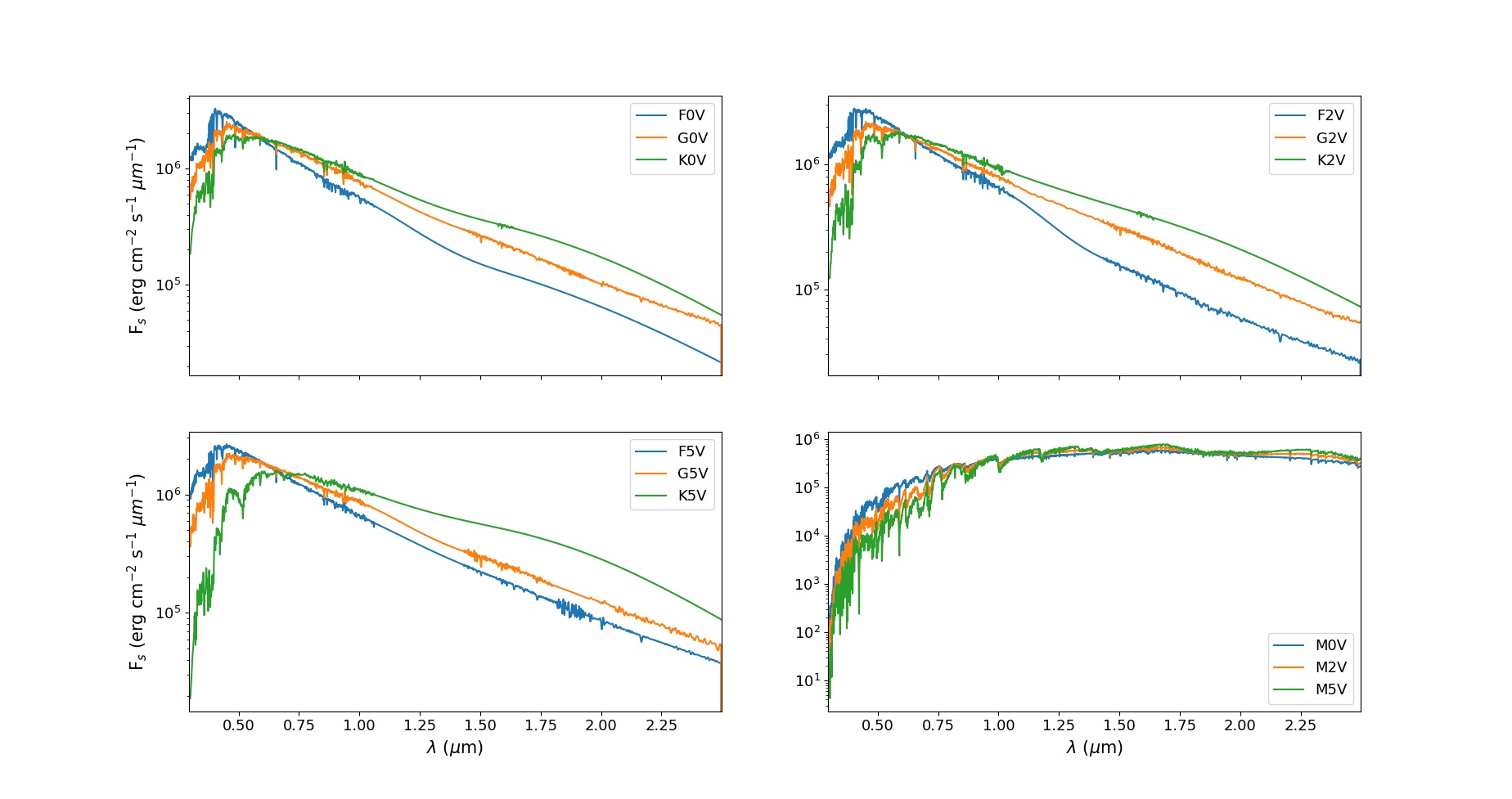}
    \caption{Input stellar flux at the surface of a habitable terrestrial planet orbiting around stars of various sub-classes of F, G, K and M spectral classes.}
    \label{fig:input1}
\end{figure}

In the present investigation, we considered terrestrial planets around stars of three sub-classes 0, 2 and 5 of F, G, K and M spectral types so that late to early stages of each spectral type are included. All the stars {were} considered to be of main sequence dwarfs of luminosity class V.  The input stellar fluxes at the surface of the planets orbiting within the habitable zone of their host stars are shown in {Figure} \ref{fig:input1}. The fluxes at the stellar surface for F, G and K spectral types {were} obtained from ESO library \citep{pickles1998stellar}. For the M spectral type, the stellar fluxes {were} obtained from PHOENIX model \citep{husser2013new} generated through publicly available code petitRADTRANS  \citep{molliere2019petitradtrans}. %These stellar spectra at the surface of the host stars are diluted appropriately to calculate the incident stellar fluxes at the surface of the exoplanets so that
The equilibrium {temperatures} of the planets {were} assumed to be same as that of the present Earth, i.e. 288 K.

\subsubsection{Reflected spectra for present Earth-like exoplanets}
We calculated the reflected spectra by solving the multiple-scattering radiative transfer equation for plane-parallel stratification (equation~\eqref{multscateqn}). To estimate the surface Bond albedo of the rocky planets, we {considered} {few} different types of surfaces with the compositions given in Table \ref{tab:albedo}.

\begin{table}
    \centering
    \begin{tabular}{|c|c|c|}
    \hline
    S.No. & Surface composition & Surface albedo\\
    \hline    
    1 & Ocean cover (100$\%$) & 0.06\\
    2 & Ocean (50$\%$), Trees and grass (50$\%$) & 0.1\\
    3 & Present Earth-like & 0.14\\
    4 & Prebiotic Earth-like & 0.16\\
    5 & Ocean (83$\%$) and snow (17$\%$) & 0.2\\
    6 & No solid or liquid surface & 0\\
    \hline
    \end{tabular}
    \caption{Surface Bond albedo for various surface compositions considered in our calculations for the modeled reflected spectra.}
    \label{tab:albedo}
\end{table}

\noindent The surface composition of the present Earth is 70$\%$ ocean, 2$\%$ coast and 28$\%$ {land, which} is divided into 30$\%$ grass, 30$\%$ trees, 9$\%$ granite, 9$\%$ basalt, 15$\%$ snow and 7$\%$ sand \citep{kaltenegger2007spectral}. And the surface composition for prebiotic Earth is 70$\%$ ocean, 2$\%$ coast and 28$\%$ land. The land surface consists of 35$\%$ basalt, 40$\%$ granite, 15$\%$ snow and 10$\%$ sand with no land vegetation \citep{kaltenegger2007spectral, rugheimer2018spectra}. {In the sixth scenario, no solid or liquid surface exists, which means that the atmosphere of the planet is so optically thick, that the incoming stellar radiation gets reflected only from the atmosphere and it does not reach up to the surface. Hence, in this case, surface albedo does not affect the reflected spectra or the geometric albedo. Zero surface albedo may also mean the gaseous planets, which is beyond the scope of this work.}

\begin{figure}
\centering
\includegraphics[width=\linewidth]{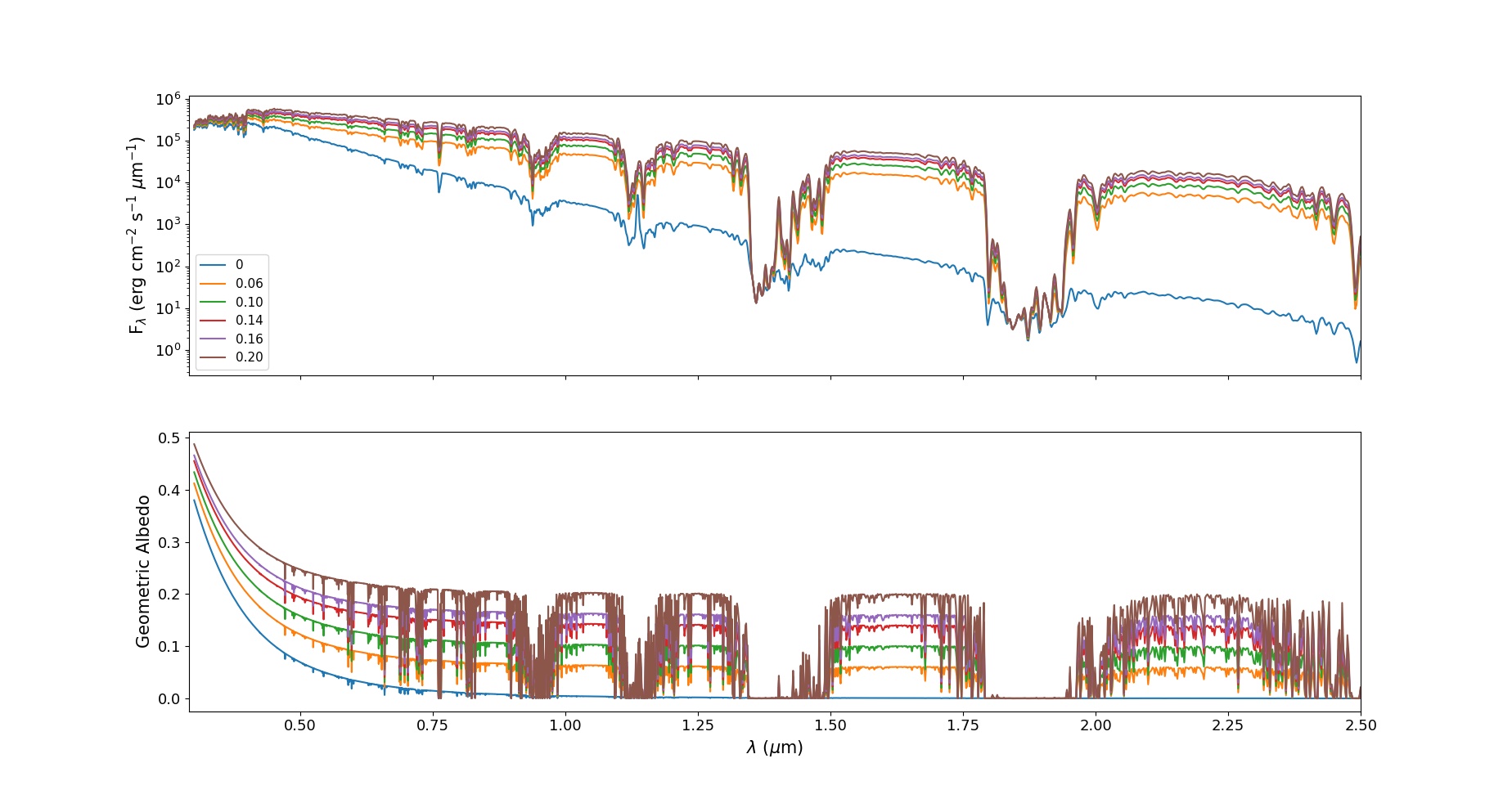}
\caption{(a) Reflected spectra for present {Earth-like} exoplanets orbiting around solar type star for different surface compositions (or different surface albedo). Blue line represents the spectra with zero surface albedo, orange line is for 100$\%$ ocean cover (surface albedo = 0.06), green line is for 50$\%$ ocean cover and the remaining 50$\%$ covered with trees and grass (0.1), red line is for present Earth-like surface composition (0.14), purple line is for prebiotic Earth-like surface composition (0.16) and brown line represents the spectra for 83$\%$ ocean and the remaining is snow (0.2). (b) Geometric albedo for the same.}
\label{fig:ref1}
\end{figure}

We calculated the surface Bond albedo by weighted sum of all the components' albedo. And the weight factors are the respective fractions of the planetary surface coverage. The reflected spectra for present Earth-like exoplanets orbiting around solar type of star for different surface albedo are shown in {Figure} \ref{fig:ref1}.  Reflected flux increases with the increase in the surface albedo and it is steeper than the input stellar flux because of Rayleigh scattering.
The effect of surface albedo on the geometric albedo for the present Earth-like exoplanets is shown in {Figure} \ref{fig:ref1}b. We can see that geometric albedo increases with the increase in surface albedo. However, it decreases with {the wavelength} because Rayleigh scattering {is not} significant at longer wavelength region.

\begin{figure}
    \includegraphics[width=\linewidth]{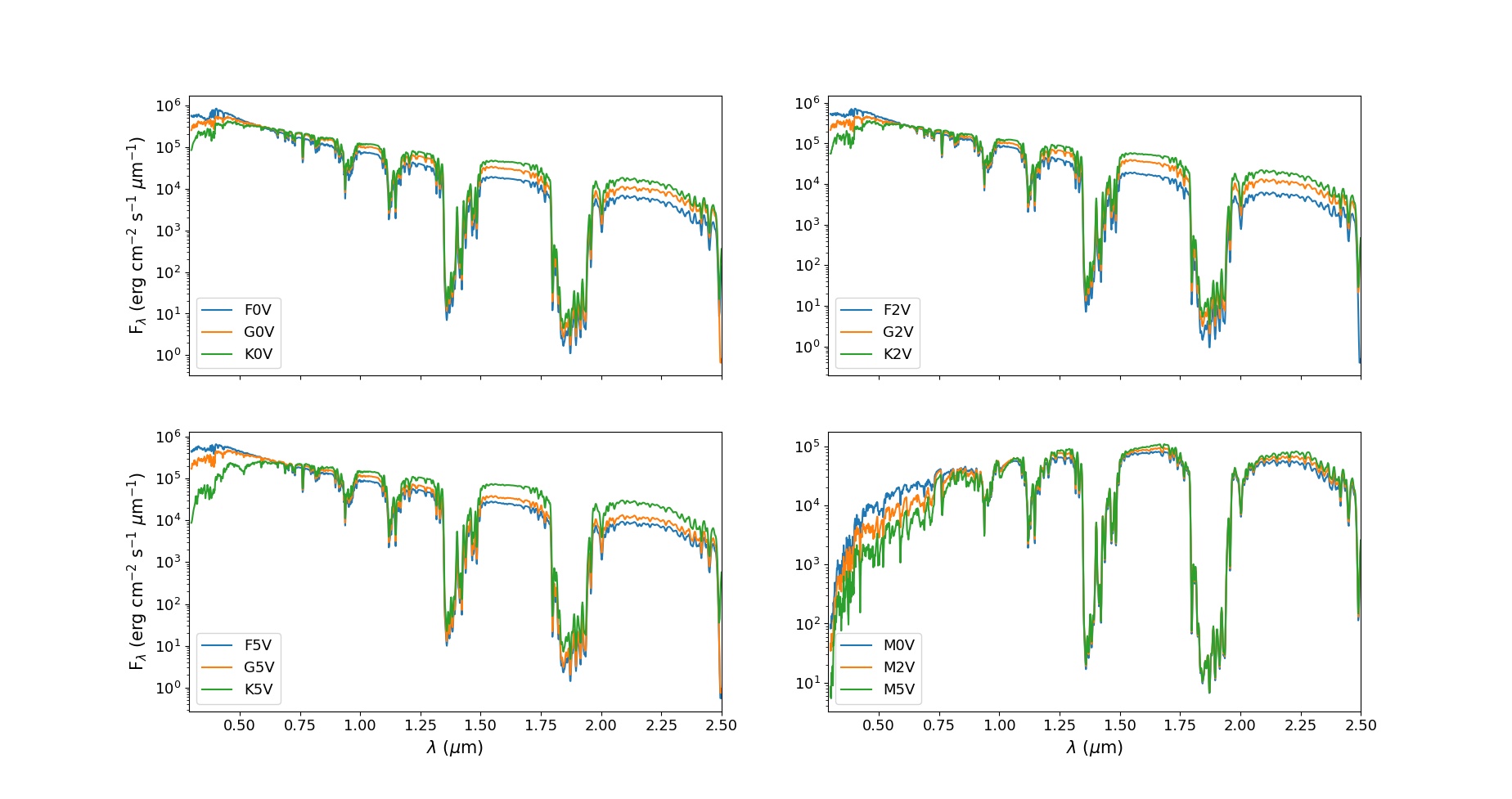}
    \caption{Reflected spectra for present {Earth-like} exoplanets orbiting around stars of spectral types F, G, K and M.}
    \label{fig:present_ref}
\end{figure}

The reflected spectra for the present Earth-like exoplanets orbiting around {stars of} F, G, K and M spectral types are shown in {Figure} \ref{fig:present_ref}. The absorption lines of {H$_2$O (0.72\,$\mu$m, 0.82\,$\mu$m, 0.94\,$\mu$m, 1.10\,$\mu$m and 1.87\,$\mu$m), O$_2$ (0.63\,$\mu$m, 0.69\,$\mu$m, 0.76\,$\mu$m) and CH$_4$ (1.60\,$\mu$m)} are also shown in this figure. The flux decreases with the increase in the wavelength in the infrared region. This is because of two reasons: firstly, the input stellar flux also decreases with the increase in wavelength in infrared and secondly Rayleigh scattering dominates in the shorter wavelength region \citep{ityaksov2008deep}. The reflected spectra has the planetary atmospheric features as well as the stellar atmospheric features.

In the present study, we {ignored} the effect of strong stellar ultra-violet irradiation that may alter the planetary environment by dissociating water molecules and energy limited hydrogen loss {\citep{sanz2011estimation, sengupta2016upper}}. Presence of sufficient initial water content at the planetary surface may still avoid the planet to become parched under such situation. {However, since most of the planets in the habitable zone of M dwarfs are tidally locked, the presence of an Earth-like planet is rare \citep{martinez2019exomoons}.}

\begin{figure}
\centering
\includegraphics[width=\linewidth]{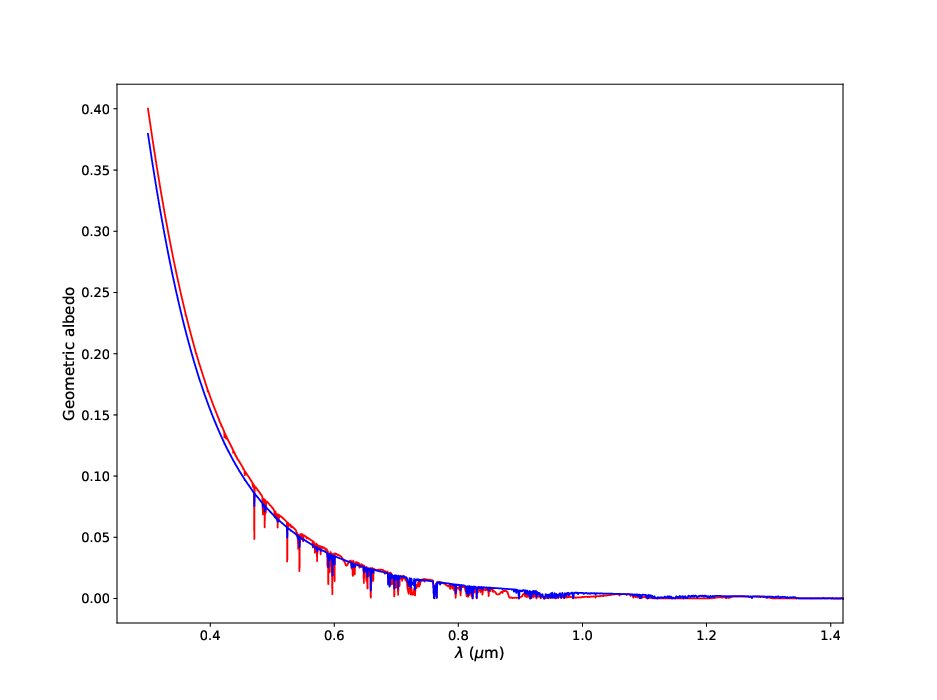}
\caption{Geometric albedo for present Earth-like exoplanets with Earth-like atmospheric composition (blue) versus  geometric albedo with increased abundances of {greenhouse} gases (red). }
\label{fig:albedo_green}
\end{figure}

In order to investigate the effect of various {greenhouse} gases on the geometric albedo, %that determines the planetary reflected flux,
we increased the abundance of CO$_2$ by {two} orders in magnitude, CH$_4$ by {four} orders in magnitude and H$_2$O by {one} order in magnitude. This increase is compensated by altering the abundance of N$_2$. The geometric albedo of the present Earth-like exoplanets with increased abundances of atmospheric greenhouse gases is presented in {Figure}~\ref{fig:albedo_green}. We found that the geometric albedo increases slightly in the shorter wavelength region because of the increase in Rayleigh scattering. However, {the scattering} could have drastic effect in the thermal re-emission at the near and far infrared wavelength region and hence in determining the {surface temperature} of the planet by an increased {greenhouse} effect.  

\subsubsection{Early Earth-like exoplanets}

\begin{figure}
    \includegraphics[width=\linewidth]{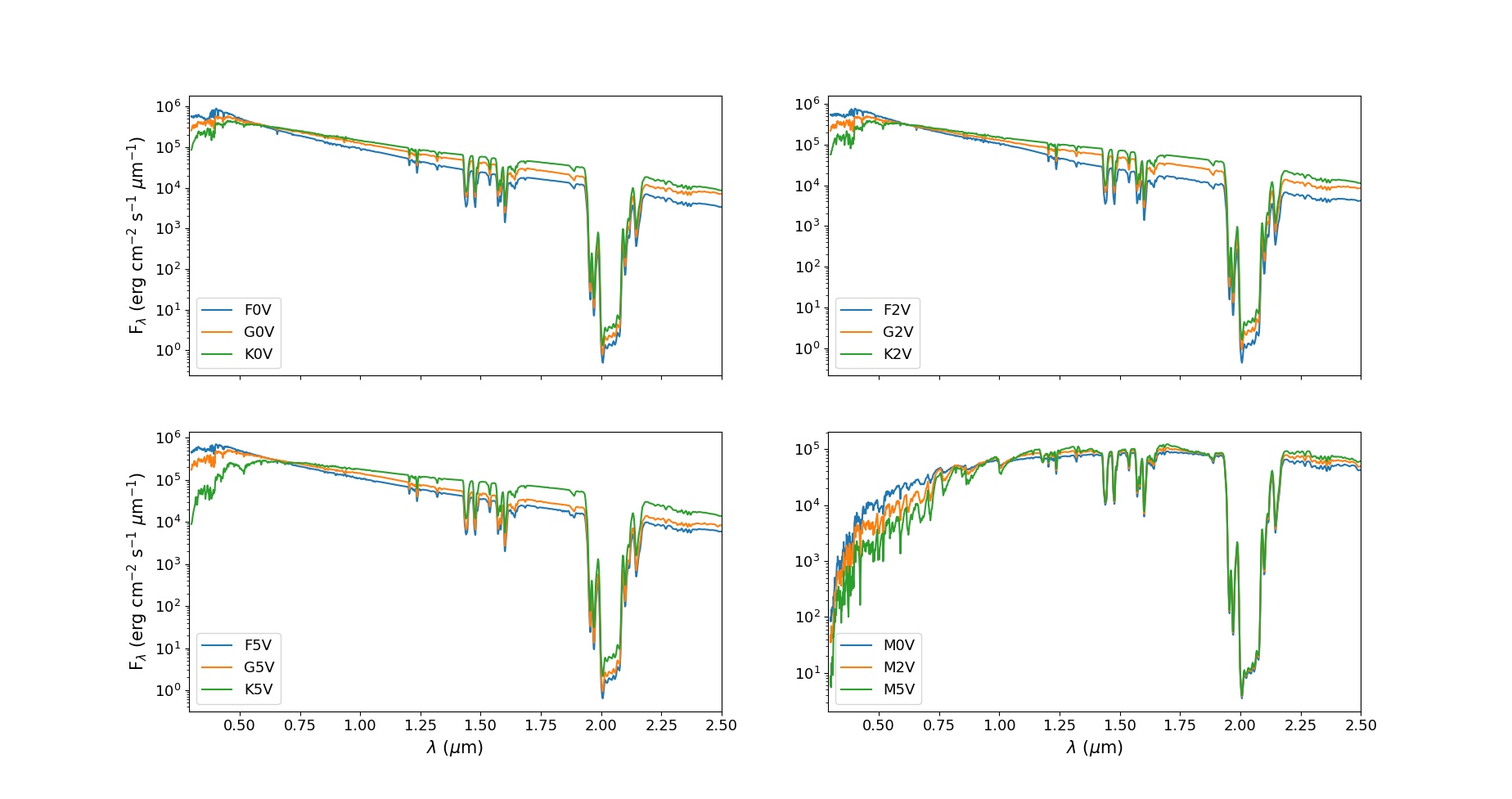}
    \caption{Same as Figure \ref{fig:present_ref} but for early {Earth-like} exoplanets.}
    \label{fig:early_ref}
\end{figure}

\begin{figure}
    \centering
    \includegraphics[width=\linewidth]{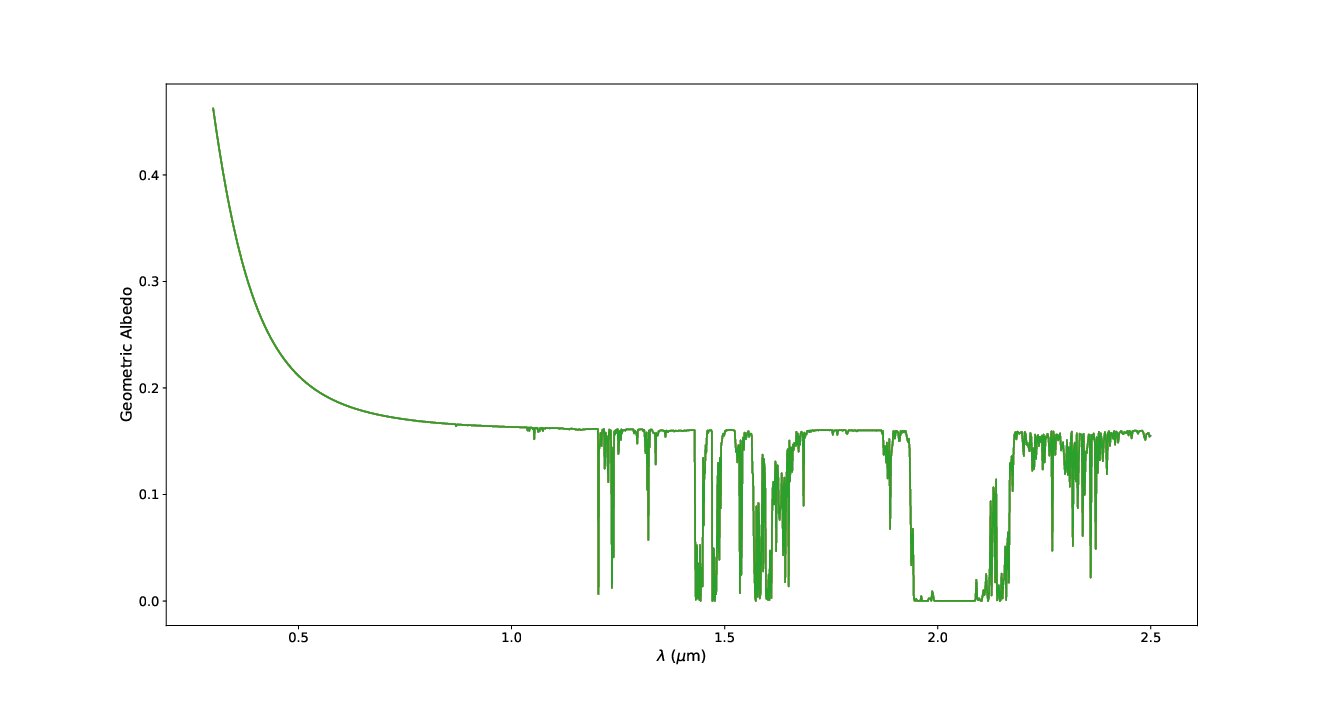}
    \caption{Geometric albedo for early Earth-like planets for surface for surface albedo = 0.16.}
    \label{fig:early_zen_alb}
\end{figure}

The reflected spectra for the prebiotic Earth orbiting around stars of F, G, K and M spectral types are presented in the Figure \ref{fig:early_ref}. And the geometric albedo (for surface albedo 0.16) is presented in Figure \ref{fig:early_zen_alb}.
We see very less absorption lines because only N$_2$, CO$_2$ and CH$_4$ {were} considered in the atmospheric composition for the prebiotic Earth. The absorption lines of CO$_2$ {(1.4\,$\mu$m, 1.6\,$\mu$m and 2\,$\mu$m) and CH$_4$ (1.66\,$\mu$m)} can be seen. The overall nature of spectra remains the same as that for the modern Earth {case}.

\begin{figure}
\centering
\includegraphics[width=\linewidth]{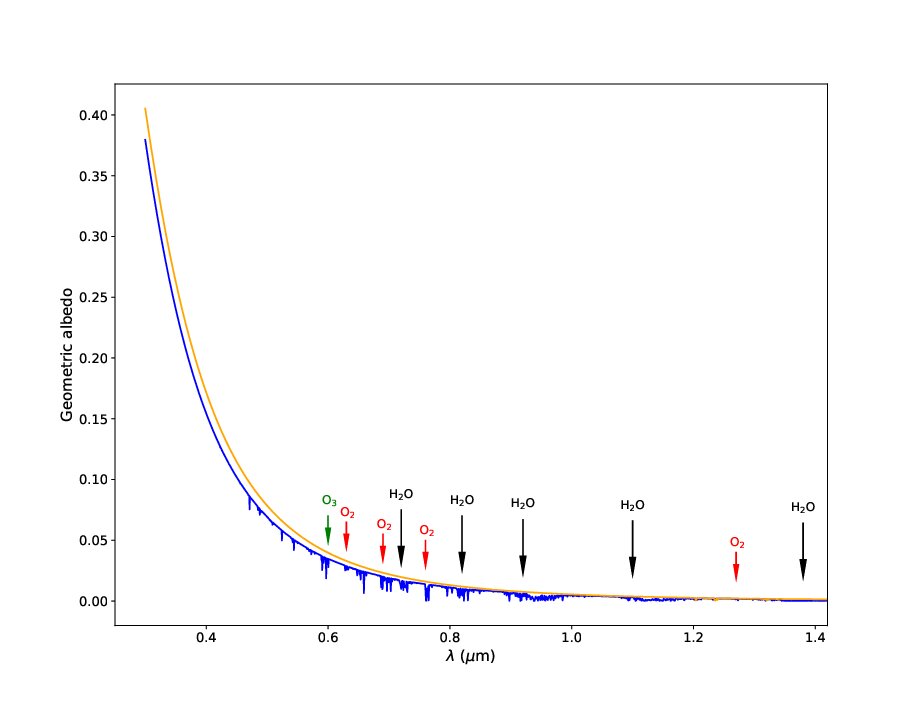}
\caption{Geometric albedo for present (blue) and prebiotic (orange) Earth-like exoplanets for zero surface albedo.}
\label{fig:albedo_modern_early}
\end{figure}

A comparison between the geometric albedo for present and prebiotic Earth with zero surface albedo is shown in the {Figure} \ref{fig:albedo_modern_early}. Prebiotic Earth-like exoplanets scatter more starlight as compared to the present Earth-like exoplanets because of greater abundances of greenhouse gases (mainly CO$_2$). The absorption lines for the present Earth are also shown in this figure.

\subsection{Reflected spectra of known terrestrial exoplanets}\label{sec:case_studies}
We also present the reflected spectra for some of the well known habitable planets such as Kepler-442b, Kepler-62e, Kepler-22b, TOI-700d, Kepler-1649c, Teegarden's {Star} b, Proxima Centauri b, {TRAPPIST}-1d and {TRAPPIST}-1e.
These planets orbit {stars of} G, K and M spectral types. Their radii are in the range of 0.7 R$_\oplus$ and 2.4 R$_\oplus$. These planets lie in the habitable planets catalog in \cite{hill2022catalog}. Although very little is known about their atmospheres at present, we expect them to have Earth-like atmospheric compositions with favourable temperature due to greenhouse effect.
 {The input stellar flux at the surface of Kepler-442b, Kepler-62e and Kepler-22b were calculated by taking the stellar flux from \cite{pickles1998stellar}. We used PHOENIX model spectra for the cases of TOI-700d, Kepler-1649c and Teegarden's {Star} b. For Proxima Centauri b, the stellar flux is taken from \citep{lin2020high} and for TRAPPIST-1d and TRAPPIST-1e, we used the spectra from \cite{burgasser2015brown}.}

 {Their equilibrium temperature T$_{\rm{eq}}$ can be derived from the relationship given in equation \ref{equilib} \citep{seager2010exoplanet}. The temperature at the bottom of the atmosphere (or surface temperature) with greenhouse effect is given by equation \ref{green} \citep{de2015planetary}.}
\begin{equation}\label{equilib}
    T_{\rm{eq}}^4 = (1-A) \frac{R_s}{2a}^2 T_{\rm{eff}}^4 
\end{equation}
           
\begin{equation}\label{green}
    T_{\rm{surf}}^4 = T_{\rm{eq}}^4 (1 + \frac{3}{4}\tau_{\rm g}) 
\end{equation}

 {In equations \ref{equilib} and \ref{green}, A is the Bond albedo, R$_s$ is the radius of the host star, $a$ is the orbital distance, {T$_{\rm{eff}}$} is the effective temperature of the host star, {T$_{\rm{surf}}$} is the temperature at the surface of the planet with greenhouse effect and $\tau_g$ is the optical depth of the atmosphere at infrared wavelengths. We assumed it to be same as that for the Earth, i.e. $\sim$ 0.83. The surface temperature should not be less than 273\,K for the planet to be habitable (T$_{\rm{surf,min}}$ $\approx$ 273\,K). And from equation \ref{green}, the minimum equilibrium temperature 
or the temperature at the top of the atmosphere (T$_{\rm{eq,min}}$) is about 242\,K.}

\subsubsection{Kepler-442b}
It is an Earth-like exoplanet orbiting its host star (K5V) within the habitable zone and about {366\,pc} away from the Earth. It is among all the detected rocky planets that is most similar to the Earth and has a very high habitability index value \citep{torres2015validation, kane2016catalog, rodriguez2017statistical}. {This planet receives an incident stellar flux that is 0.9 times of the flux received by the Earth \citep{torres2015validation, armstrong2016host, rodriguez2017statistical, barbato2018revised}.} {It} is a promising candidate for search of biosignatures as K-type of stars maintain favourable circumstellar conditions for habitability \citep{cuntz2016exobiology}. {Its density is very similar to the Earth and mean surface gravity is $\sim$ 12.5 m/s$^2$, slightly higher than that of the Earth}. According to \cite{arney2019k}, K-type stars present an advantage for the detectability of biosignatures. {One of the reasons is that K dwarfs offer extended photochemical lifetime of methane as compared to G types stars. And the other reason is better signal-to-noise ratio (S/N) of {K dwarfs} than {G dwarfs}, due to which oxygen and methane can be strongly observed.} {We calculate the $T$-$P$ profile by the following method:}

\begin{enumerate}
\item For tropospheric region ( {up to} {11\,km}), T = -mh + {T$_{\rm{surf}}$} where m is the adiabatic lapse rate.\\
 {T$_{\rm{surf,min}}$} $\approx$ 273\,K;   For h=11\,km, T $\approx$ 242\,K; m = 2.83\,K/km\\
 {T$_{\rm{surf,max}}$} $\approx$ 290.1\,K;   For h=11\,km, T $\approx$ 257\,K;  m = 3\,K/km
\begin{equation}
    T_1 = - 2.83 h + 273;\\
    T_2 = -3 h + 290.1
\end{equation}

\item For Stratospheric region and above,
T$_1$ $\approx$ 242\,K; T$_2$ $\approx$ 257\,K.

\end{enumerate}

\begin{figure}
    \centering
    \includegraphics[width=\linewidth]{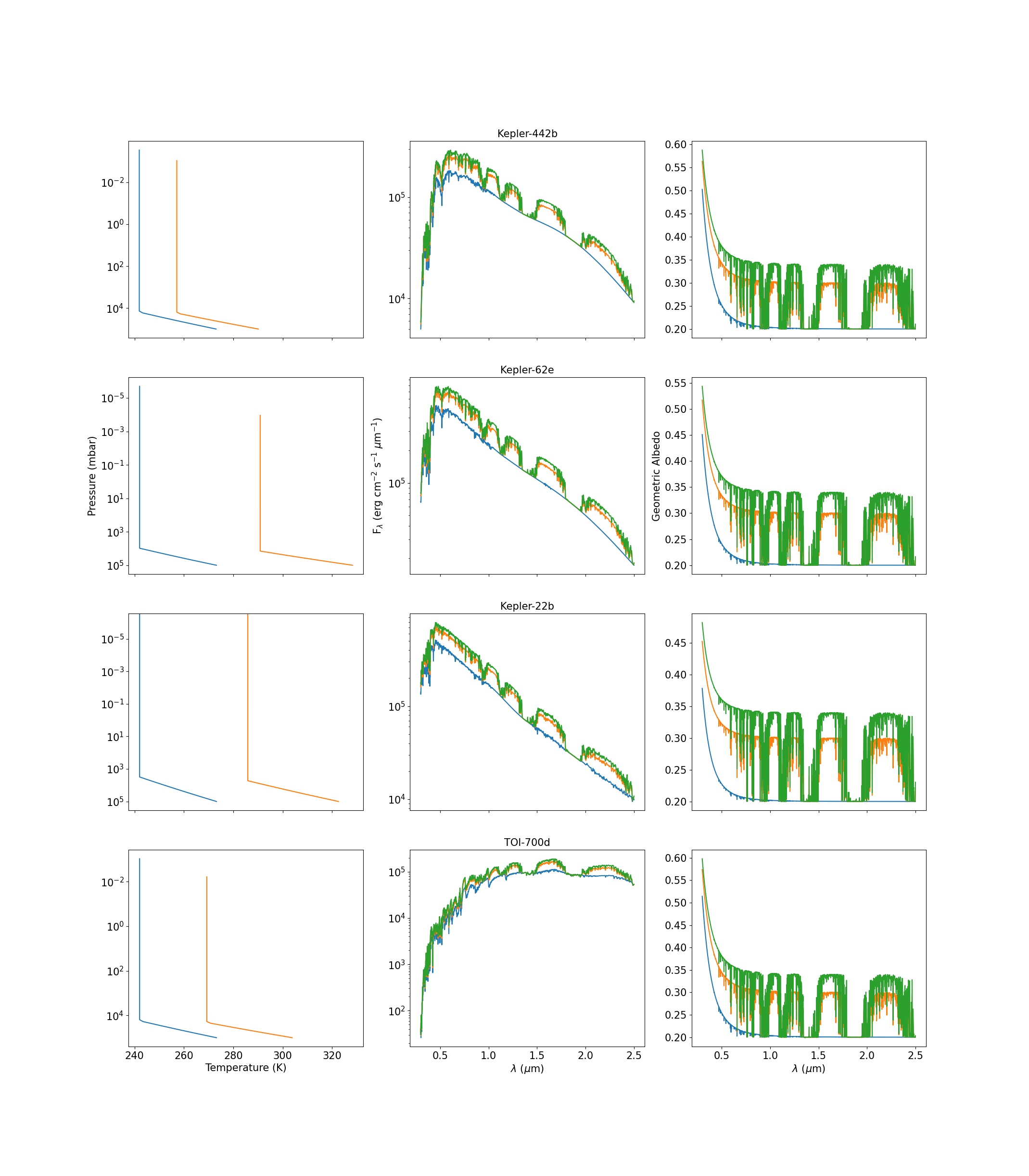}
    \caption{(a) Possible range of $T$-$P$ profile; Effect of the surface albedo on (b) the reflected spectra; and (c) geometric albedo for {Kepler-442b, Kepler-62e, Kepler-22b and TOI-700d}. Green curve is for the Earth-like surface composition (surface albedo = 0.14), orange for 50\% ocean and 50\% land consisting of trees and grass only (0.1) and blue for zero surface albedo (i.e., no solid/liquid surface)}
    \label{fig:kep-442b}
\end{figure}

 {Similarly, we calculated the $T$-$P$ profile for all the other planets by calculating their adiabatic lapse rates.}
The possible range of $T$-$P$ profile for Kepler-442b is shown in {Figure} \ref{fig:kep-442b} and the temperature can lie anywhere in this range. The maximum value of Bond albedo (A$_{\rm{max}}$), temperature at the top of the atmosphere ( {T$_{\rm{eq}}$}) and temperature at the bottom of the atmosphere including green-house effect ( {T$_{\rm{surf}}$}) {were} calculated in the same ways and are shown in Table \ref{table2}.
The atmospheric abundance {was} assumed to be the same as that of the Earth and {shown} in {Table} \ref{table}. We {calculated} the reflected spectra for the two $T$-$P$ profiles and {found} that the spectra does not alter with the variation in $T$-$P$ profile within the given range. The reflected spectra for the planet Kepler-442b for {various} surface compositions i.e. different surface albedos is shown in the {Figure} \ref{fig:kep-442b}.

 {Figure} \ref{fig:kep-442b} also shows the geometric albedo of Kepler-442b for different surface compositions of the planet. {We note that the geometric albedo increases significantly with the increase in the surface Bond albedo or 
we can say that the surface albedo considerably affects the geometric albedo. This is because the surface also contributes in the total reflectivity of the planet. For example, for the zero surface albedo case, the geometric albedo is the least. And it is maximum for the present Earth-like surface components (0.14 surface albedo).}

\begin{table}
\centering

\begin{tabular}{|c|c|c|c|c|c|c|c|}
\hline
Planet & T$_{\rm{eq,max}}$ & T$_{\rm{surf,max}}$ & A$_{\rm{Bond,max}}$ & R$_{\rm p}^{[1]}$ & M$_{\rm p}^{[1]}$ & a$^{[1]}$ & ESI$^{[2]}$ \\
 & (K)      & (K)        &          & (R$_\oplus$) & (M$_\oplus$) & (au) & \\
\hline
Kepler-442b & 257 & 290 & 0.216 & 1.34 & 2.36 & 0.409 & 0.84\\
Kepler-62e & 291 & 328 & 0.52 & 1.61 & 36 & 0.427 & --\\
Kepler-22b & 286 & 323 & 0.486 & 2.33 & 36 & 0.849 & --\\
TOI-700d & 269 & 304 & 0.347 & 1.144 & 1.57 & 0.1633 & 0.93\\
Kepler-1649c & 296 & 334 & 0.55 & 1.06 & 1.2 & 0.0649 & 0.90\\
Teegarden b & 289 & 326 & 0.51 & 1.02$^*$ & 1.05 & 0.0252 & 0.95\\
Proxima b & 258 & 292 & 0.2289 & 1.08$^*$ & 1.27 & 0.0485 & 0.87\\
TRAPPIST-1d & 286 & 323 & 0.49 & 0.788 & 0.388 & 0.0223 & 0.90\\
TRAPPIST-1e & 250 & 282 & 0.12 & 0.92 & 0.692 & 0.0292 & 0.85\\
\hline
\end{tabular}
\caption{ {T$_{\rm{eq,max}}$} is the maximum temperature at the top of the atmosphere of the planet and {T$_{\rm{surf,max}}$} is the maximum temperature at the bottom of the atmosphere after considering greenhouse effect. A$_{\rm{max}}$ is the maximum possible Bond albedo, ESI is the Earth similarity index, R$_p$ and M$_p$ are the radius and mass of the planet and a is the orbital {separation}.
[1]\url{https://exoplanets.nasa.gov/exoplanet-catalog/} [2]\url{https://phl.upr.edu/projects/earth-similarity-index-esi}\\
*estimate value}
\label{table2}
\end{table}

\subsubsection{Kepler-62e}

Kepler-62e also orbits within the classical habitable zone of the host star (K2V) and the orbital period is about 122 days \citep{borucki2013kepler, kaltenegger2013water, torres2015validation, armstrong2016host, kane2016catalog}. The possible $T$-$P$ profile {was} calculated in the same way as in the case of Kepler-442b and is presented in {Figure} \ref{fig:kep-442b}. The temperature and pressure can be anywhere between these limits.

The reflected spectra and the geometric albedo for Kepler-62e are also shown in {Figure} \ref{fig:kep-442b} for various surface compositions. It is highest for Earth-like surface composition (surface albedo 0.14) and lowest for no surface albedo at all. The green curve is for 50$\%$ ocean cover and the remaining covered with trees and grass. As the ocean cover is reduced from 70$\%$ to 50$\%$ by increasing the land cover, the geometric albedo decreases.

\subsubsection{Kepler-22b}

Kepler-22b is a super-Earth orbiting a G5V star {, which} is about {194.7\,pc} away from Earth. This planet is also orbiting within the habitable zone of the host star \citep{borucki2012kepler, neubauer2012life, torres2015validation, kane2016catalog}. It is the first detected Earth-like exoplanet in the habitable zone of a solar-type star.

The atmospheric $T$-$P$ profile used to calculate the reflected spectra of the planet is shown in {Figure} \ref{fig:kep-442b}. It is also calculated by assuming the atmosphere of the planet in hydrostatic equilibrium and considering greenhouse effect.

The effect of the surface Bond albedo (derived from the surface compositions) on the reflected spectra and the variation of the geometric albedo are also shown in {Figure} \ref{fig:kep-442b}.

\subsubsection{TOI-700d}

It is TESS’s first Earth-size exoplanet {, which} lies in the habitable zone of its host star TOI-700 ( {M dwarf}). The planet is expected to be tidally locked as its eccentricity is close to zero \citep{gilbert2020first, rodriguez2020first, suissa2020first, kaltenegger2021around}. It receives about 86$\%$ of the insolation that the Earth receives \citep{gilbert2020first}. The possible range of the {atmospheric} $T$-$P$ profile for this planet is shown in {Figure} \ref{fig:kep-442b}. The reflected spectra and the geometric albedo are also shown in {this figure}.

\subsubsection{Kepler-1649c}
\begin{figure}
    \centering
    \includegraphics[width=0.94\linewidth]{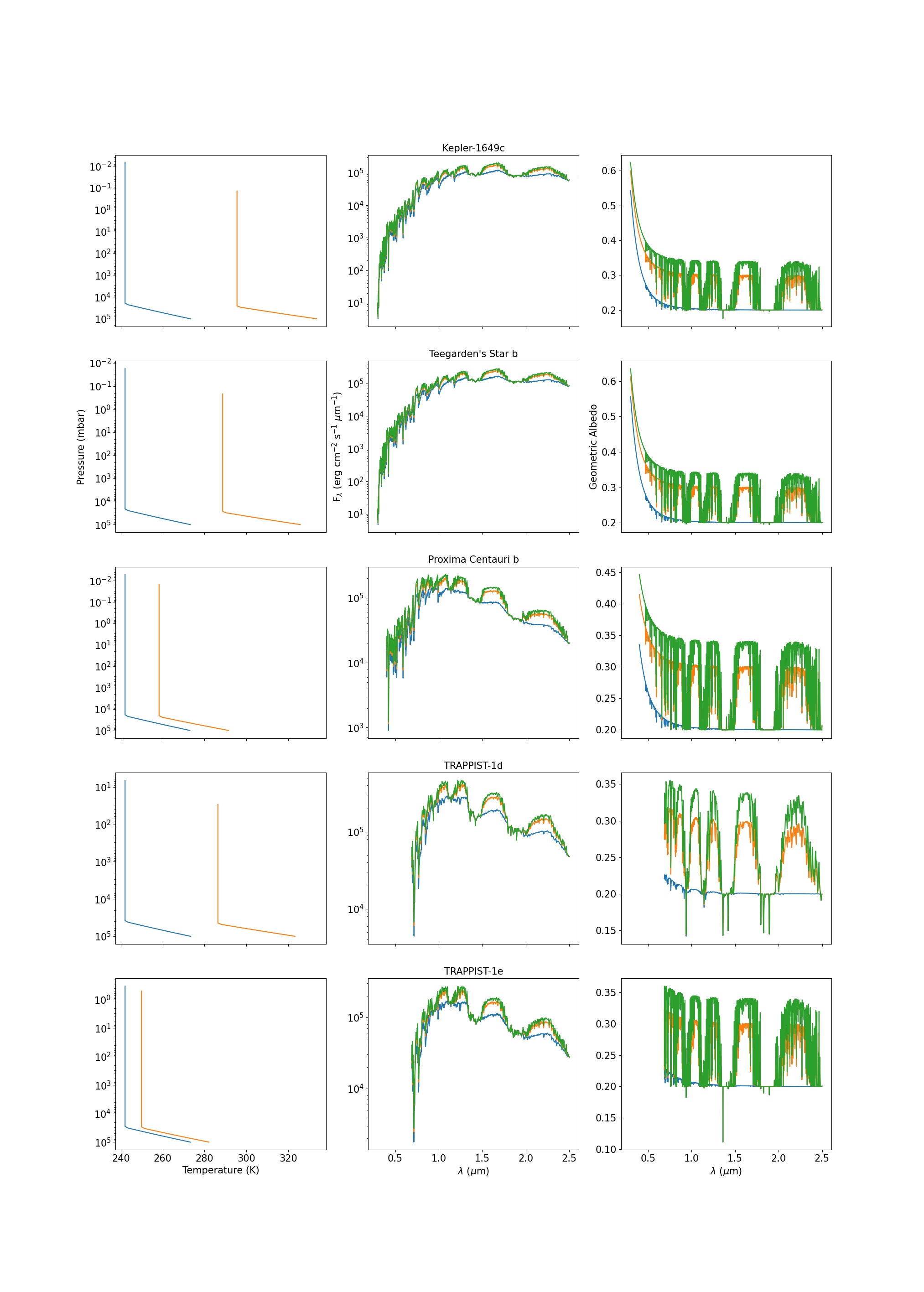}
    \caption{Same as {Figure} \ref{fig:kep-442b} but for {Kepler-1649c, Teegarden's Star b, Proxima Centauri b, TRAPPIST-1d and TRAPPIST-1e}.}
    \label{fig:panel_2}
\end{figure}

This is an Earth-size planet lying in the habitable zone of its host star {, which} is {of M5V spectral type}. It is located at a distance of about {92\,pc} from the Earth \citep{vanderburg2020habitable, kane2020eccentricity, gvalaniquestionnaire}. The $T$-$P$ profile, the reflected spectra and the geometric albedo are presented in {Figure} \ref{fig:panel_2}.

\subsubsection{Teegarden's Star b}
Teegarden's Star was discovered by \cite{teegarden2003discovery} and it is at a  distance of {3.831\,pc} and of spectral type M7V \citep{alonso2015carmenes}. {It has two planets Teegarden's Star b and c. Both of them are super Earths but Teegarden b is} the most Earth-like planet or maximum ESI value (see Table \ref{table2}), discovered till now \citep{wandel2019habitability, zechmeister2019carmenes}. This planet lies within the habitable zone and it is tidally locked. The T-P profile range, reflected spectra and the geometric albedo are presented in {Figure} \ref{fig:panel_2}.

\subsubsection{Proxima Centauri b}

Proxima Centauri b is a rocky planet that orbits within the habitable zone of our nearest neighbour Proxima Centauri (M5V), which receives about 65$\%$ of the total flux that our Earth receives from the Sun \citep{anglada2016terrestrial, garraffo2016space, turbet2016habitability, ribas2017full, meadows2018habitability, lin2020high, galuzzo2021three}.

We {modeled} the reflected spectra of Proxima Centauri b by using the stellar flux presented by \cite{lin2020high}. The $T$-$P$ profile (derived in the same way) for the atmosphere of Proxima Centaui b is shown in {Figure} \ref{fig:panel_2} where a range is given. The reflected spectra and the geometric albedo for Proxima Centauri b {are} shown in {Figure} \ref{fig:panel_2} for various surface Bond albedo determined by different surface compositions. Here also, it is maximum for Earth-like surface composition and minimum for no surface albedo.

\subsubsection{ {TRAPPIST-1d and e}}
 {TRAPPIST}-1 is another {M dwarf} of spectral type M8V {, which} is about {12\,pc} away from us, hosts seven rocky planets out of {which,} three are in the habitable zone of the star \citep{gillon2016temperate, burgasser2017age, gillon2017seven, o2019lessons, lin2020high}.
 {TRAPPIST}-1e is most likely to have habitable surface conditions, as it receives about $\sim$ 66$\%$ of stellar radiation that the Earth receives from the Sun and needs very little greenhouse effect to have a surface temperature such that water can exist in liquid state \citep{kopparapu2013habitable, wolf2017assessing, wolf2018erratum, fauchez2020trappist}. Also, {TRAPPIST}-1e is quite similar in size to the Earth. {On the other hand, TRAPPIST-1d has a very high ESI value of 0.9 (see Table \ref{table2}). So it becomes important to model both the planets.}

The $T$-$P$ profile, the reflected spectra and the geometric albedo are presented in {Figure} \ref{fig:panel_2} for both the cases.
The reflected spectra and the geometric albedo {were} calculated for various surface materials. 
As the surface albedo increases, the reflected flux increases because the surface also contributes to the reflected flux.
The geometric albedo is decreasing with the increase in wavelength because scattering becomes negligible at longer wavelengths. Also it decreases significantly with the decrease in the surface {Bond} albedo. For the zero surface albedo case, all the radiation is reflected only from the atmosphere.

\subsubsection{Comparisons}
\begin{figure}
    \centering
    \includegraphics[width=0.9\linewidth]{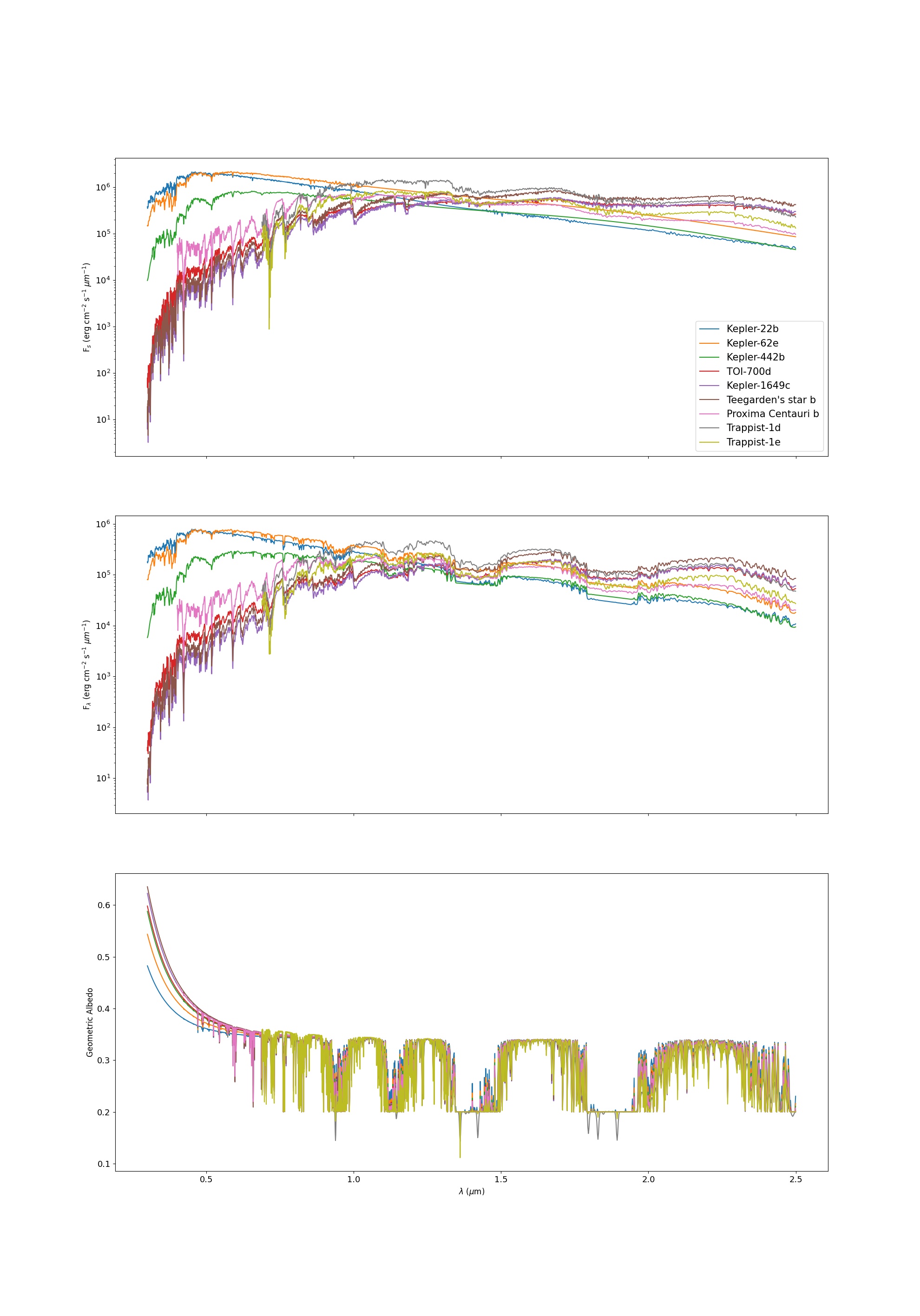}
    \caption{Comparison of the (a) input stellar flux, (b) reflected spectra and (c) geometric albedo for {the nine planets in habitable zone}.}
    \label{fig:comp}
\end{figure}

The input stellar flux at the surface of {the above planets is shown in} {Figure} \ref{fig:comp}a.
The reflected spectra for these planets for Earth-like surface albedo are shown in {Figure} \ref{fig:comp}b and the geometric albedo is shown in the {Figure} \ref{fig:comp}c. We can see that the reflected spectra follows the input stellar spectra in the visible wavelength region. 
%But the input spectra for the case of TRAPPIST-1e is higher in infrared but the reflected spectra of it becomes the least in infrared. The reflected spectra includes the spectral features of stellar as well as the planetary atmosphere while the geometric albedo contains the atmospheric features of the planet only as the stellar flux gets cancelled out in its calculation.
The geometric albedo is highest for Kepler-22b and lowest for {TRAPPIST}-1e in the infrared. But in the optical, it is highest for the case of Teegarden's {Star} b and lowest for Kepler-22b.

The geometric albedo in the optical region {is not} estimated for any of the planets around {TRAPPIST}-1 {because it is a late M dwarf whose effective temperature is about 2400K. Its blackbody spectra peak lies at around 1\,$\mu$m and thus the flux in the optical is very less in magnitude as compared to the flux in NIR}. The nature of geometric albedo depends on the absorption and scattering co-efficients of the planetary atmosphere or the $T$-$P$ profile and the atmospheric composition of the corresponding planet.

%%%%%%%%%%%%-------------transmission spectra ---------------%%%%%%%%% %  %%%%% %%%%%%%%%%%

\subsection{The Transmission Spectra}\label{sec:trans_spectra}
When an exoplanet {transits} in front of {its} host star, it blocks some of the starlight along our line of sight resulting into a reduction in the observed  stellar flux. During the transit, a fraction of the stellar radiation passes through the planetary atmosphere providing {signatures} of the {gases} present there. The stellar radiation that suffers absorption and scattering in the atmosphere of the planet is known as the transmission or transit spectra. The transmission spectra is usually presented by a wavelength dependent quantity called the transmission or transit depth. 

\subsubsection{Transit Depth}
Transit depth is the ratio between the stellar flux obtained with and without transit. It can be expressed as
\begin{eqnarray} \label{eq:transit}
    D = 1 - \frac{F_{trans}}{F_{star}}
\end{eqnarray} 
where $F_{trans}$ is the stellar flux obtained during the planetary transit epoch and $F_{star}$ is the unblocked stellar flux or the stellar flux during the out of transit  epoch. $F_{trans}$ can be written as \citep{kempton2017exo, sengupta2020optical}:
\begin{eqnarray} \label{eq:f_trans}
    F_{trans} = [1-(\frac{R_{pl,atm}}{R_{star}})^2] F_{star} + F_{atm}
\end{eqnarray}
where, $R_{pl,atm}$ is the radius of the planet including its atmosphere and $R_{star}$ is the radius of the star, $F_{atm}$ is the stellar flux {, which} gets transmitted through the planetary atmosphere along the line of sight. Transit depth corresponds to the ratio between the planetary radius $R_{pl,atm}$ and the stellar radius $R_{star}$.

\begin{figure}
    \includegraphics[width=\linewidth]{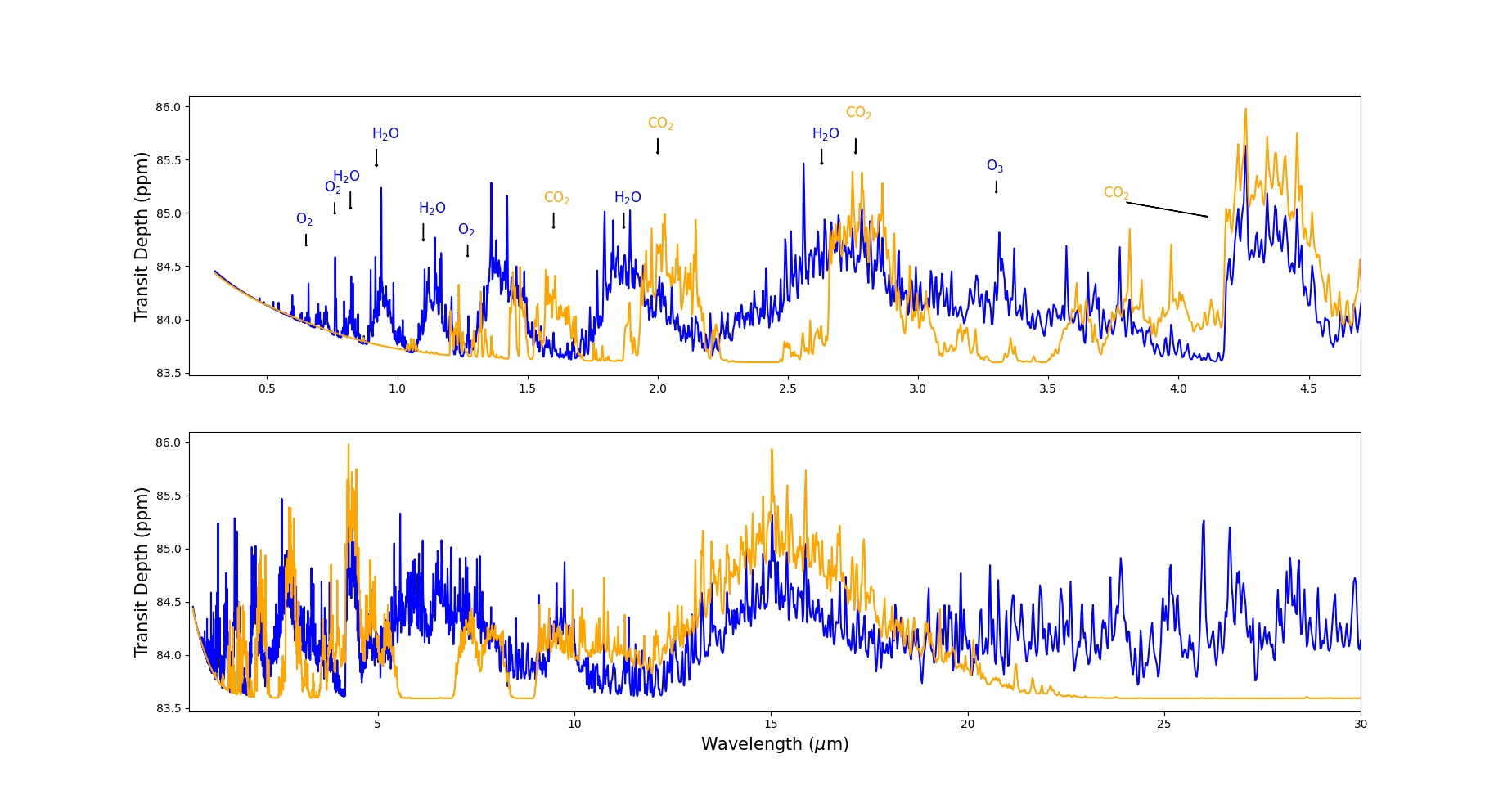}
    \caption{ Transmission spectra of present (blue) and early (orange) Earth-like exoplanets.}
    \label{fig:trans_exo}
\end{figure}

In order to calculate  $F_{atm}$, we used Beer-Bouguer-Lambert's law given by:
\begin{eqnarray} \label{eq:bbl}
    I(\lambda) = I_{0}(\lambda) e^{-\tau_\lambda/\mu_0}
\end{eqnarray}
where $I(\lambda)$ is the intensity of the transmitted stellar radiation through the planetary atmosphere and $I_0(\lambda)$ is the intensity of the incident stellar radiation on the planet. In the above equation, $\tau_\lambda$ is the optical depth along the ray path and $\mu_0$ is the cosine of the angle between the direction of the incident radiation and the normal. The expression for the optical depth along the line of sight ($\tau(\lambda,z)$) is given by:
\begin{eqnarray}
    \tau(\lambda,z) = 2\int_0^{l(z)} \chi(\lambda,z)\rho(z)dl 
\end{eqnarray}
where $\chi(\lambda,z)$ is the extinction coefficient {, which} is the sum of the absorption coefficient and the scattering coefficient, $\rho(z)$ is the density of the planetary atmosphere, $z$ is the height of the atmosphere from the planetary surface, $l$ is the distance covered by the radiation in the planetary atmosphere given by \citep{Tinetti}:
\begin{eqnarray}
    l(z) = \int dl = \sqrt{(R_p + z_{max})^2 - (R_p + z)^2}
\end{eqnarray}
where $R_p$ is the radius of the planet {, below which} the medium becomes opaque at all wavelength and $z_{max}$ is the maximum height (on the top of $R_p$) {, above which} photons {do not} suffer any absorption or scattering. We {calculated} the transmission spectra by using {\tt Exo-Transmit} package \citep{kempton2017exo}. {Figure} \ref{fig:trans_exo} shows the transmission spectra for the present and the prebiotic Earth-like exoplanets. {Figure}~\ref{fig:trans_exo}a presents the transmission spectra {up to} a wavelength of {4.5\,$\mu$m }while {Figure} \ref{fig:trans_exo}b shows the same {up to} the wavelength {30\,$\mu$m}. The transmission depth due to absorption by O$_2$, H$_2$O, CO$_2$ and O$_3$ are marked in the spectra. For the early or prebiotic Earth-like exoplanets, the absorption lines of only CO$_2$ molecules are seen.

In the transmission {spectra} of prebiotic {Earth-like exoplanets} presented in {Figure} \ref{fig:trans_exo}(a) {, which is the zoomed-in version of {Figure} \ref{fig:trans_exo}(b),} signatures of CO$_2$ can be found at {1.4\,$\mu$m, 1.6\,$\mu$m, 2.0\,$\mu$m, 2.7\,$\mu$m and at 4.3\,$\mu$m.} On the other hand, the signatures of H$_2$O at {0.72\,$\mu$m, 0.82\,$\mu$m, 0.94\,$\mu$m, 1.10\,$\mu$m, 1.87\,$\mu$m and 2.70\,$\mu$m} are clear in the transmission {spectra} of modern Earth. In the transmission {spectra} of modern Earth, signatures of O$_2$ can also be found at {0.63\,$\mu$m, 0.69\,$\mu$m, 0.76\,$\mu$m and that of CO$_2$  at 1.4\,$\mu$m, 2.7\,$\mu$m, 4.3\,$\mu$m}. The signature of  O$_3$ is visible at {3.3\,$\mu$m}. { {Figure} \ref{fig:trans_exo}(b) shows the transmission spectra for the whole wavelength region i.e. 0.3$\mu$m to 30.0$\mu$m.}

%\subsection{Transmission spectra with increased %abundances of greenhouse gases}

\begin{figure}
\centering
\includegraphics[width=\linewidth]{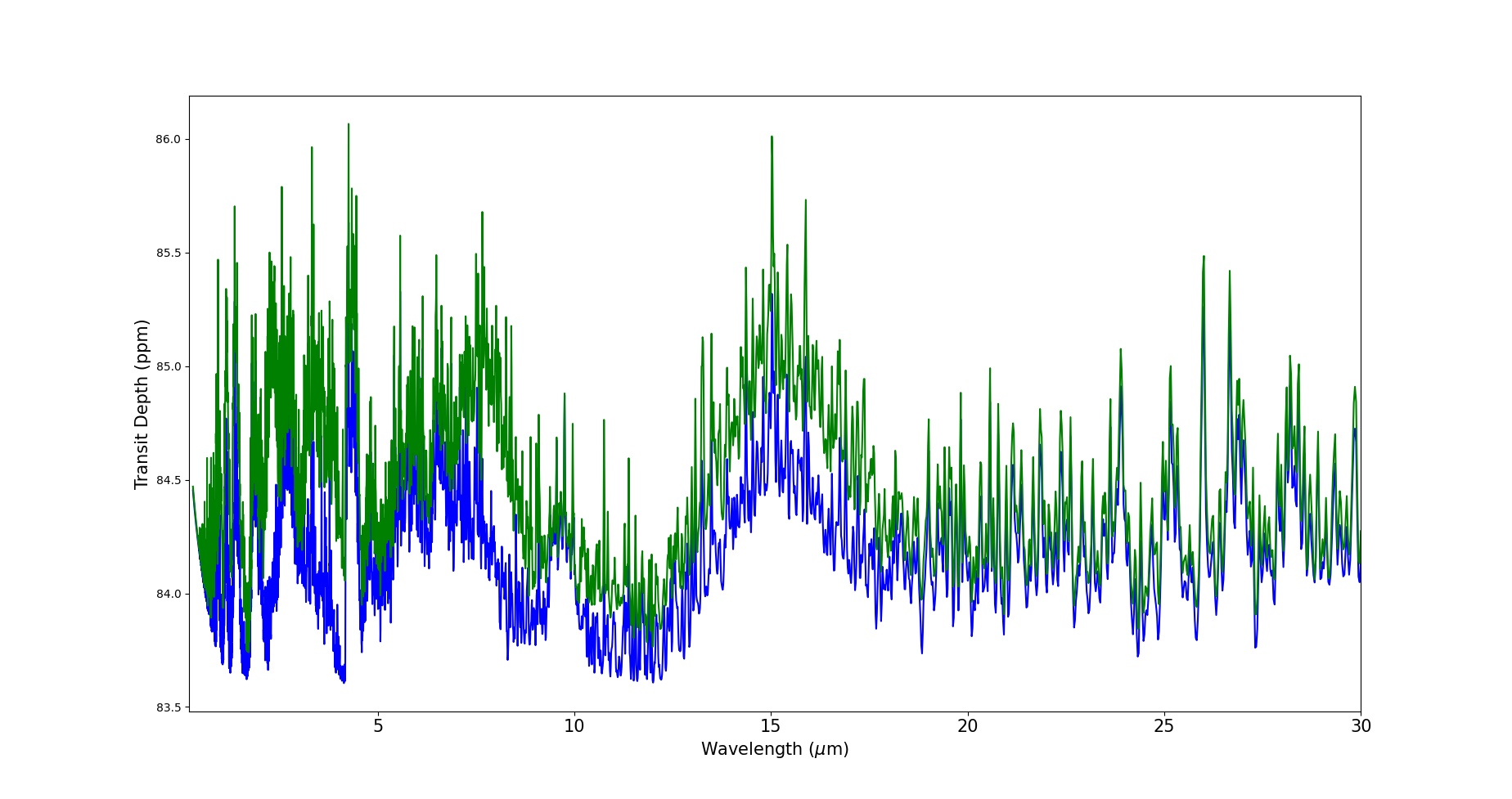}
\caption{Transmission spectra of the present Earth but with increased greenhouse gas abundance (green). For a comparison, transmission spectrum (blue) of the Earth with actual atmospheric abundance is also presented. }
\label{fig:trans_green}
\end{figure}

 {Figure} \ref{fig:trans_green} shows the transmission spectra for the increased abundance of greenhouse gases in the atmosphere of the terrestrial exoplanet. Here we notice that the transmission depth increases with the increase in the abundance of greenhouse gases in the planetary atmosphere. However, this increase in the transmission depth is found to be confined only up to a certain wavelength region, {which is again due to Rayleigh scattering}. 

\subsubsection{Effects of cloud opacity}
The observations of the various exoplanetary atmospheres indicate that the presence of clouds or hazes are common phenomenon in the planetary atmospheres \citep{kreidberg2014clouds, sing2016continuum}. This is one of the reasons for weak or no molecular feature observed in the transmission spectra of quite a few hot Jupiters {\citep{sanchez2020discriminating}}. The same situation may arise for the terrestrial exoplanets if the upper atmosphere is covered by clouds or hazes. The presence of clouds or hazes however, increases the Rayleigh scattering.

\begin{figure}
\centering
\includegraphics[width=\linewidth]{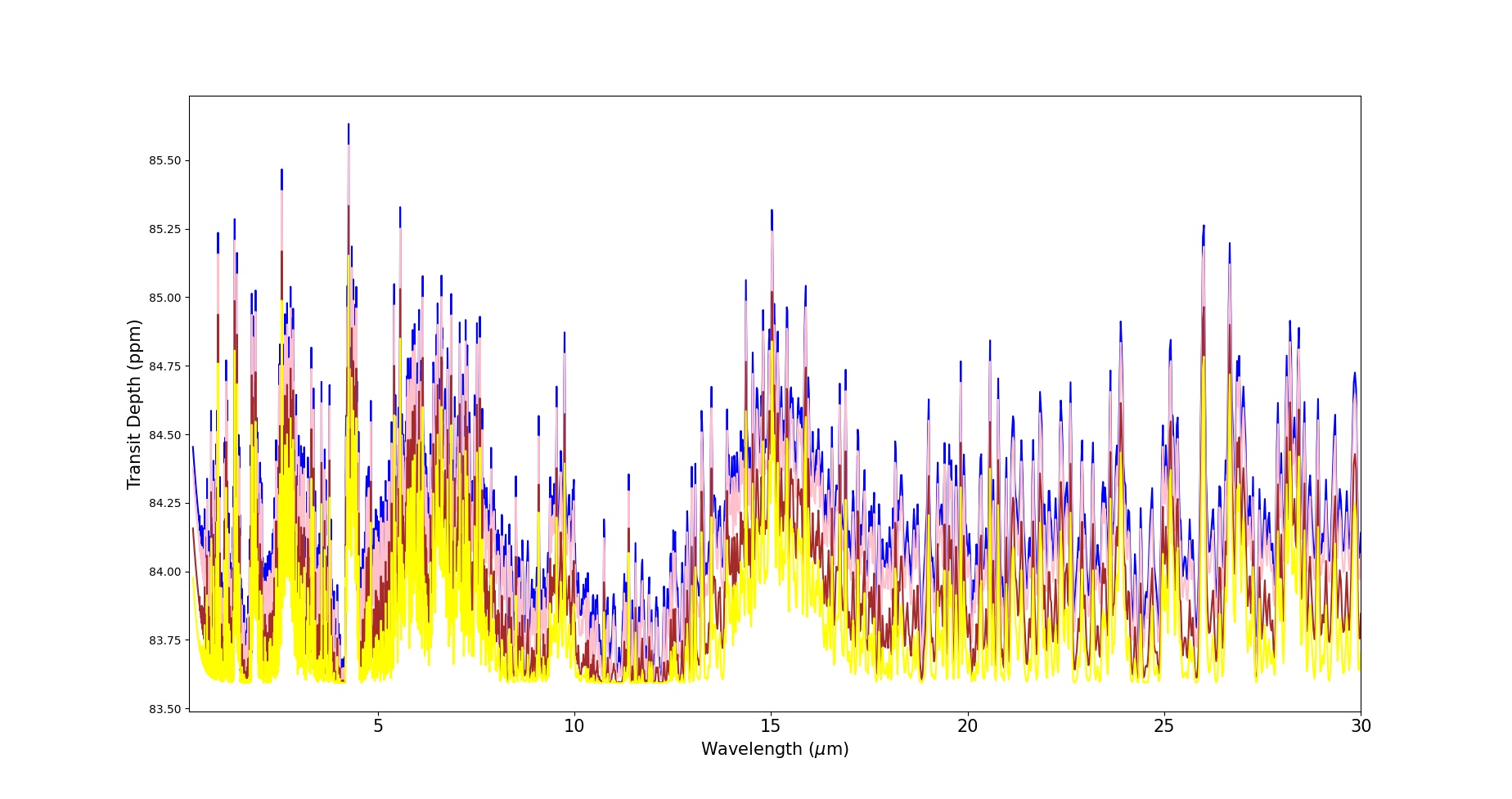}
\caption{Transit spectra for present Earth with clear sky (blue) and for 100 percent cloud coverage at three different heights i.e. 2.2 km (pink), 9.5 km (brown) and {17.0} km (yellow) from the surface of the planet.}
\label{fig:trans_clouds}
\end{figure}

For the gray cloud calculation, we selected a pressure layer in the atmosphere at which the cloud top is optically thick. We provided a threshold pressure within the pressure range of the $T$-$P$ profile and performed the radiative transfer calculations for pressures below that of the cloud deck. We used {\tt Exo-Transmit} package \citep{kempton2017exo} for the calculation of cloud optical depth.
 {Figure} \ref{fig:trans_clouds} shows the transmission spectra for clear sky and for the sky with 100$\%$ coverage of clouds at three different atmospheric heights i.e. 2.2\,km, 9.5\,km and {17.0}\,km from the surface of the planet.

%%%%%%%%%%%%%-------------conclusions and discussions------------%%%%%%%%%%%%%%%%%%%%

\section{Conclusions and Discussion}\label{sec:disc_conc}

In the first part of this paper, we presented the numerical models of reflection spectra (in visible) for both the present and prebiotic Earth-like exoplanets orbiting within the habitable zone of main sequence stars of F, G, K and M spectral types. We also {presented} the model reflected spectra for the known exoplanets {, which} are orbiting around {the stars of} G, K and M spectral types.

We found that the nature of the reflected spectra is similar to that of the incident stellar spectrum i.e., the reflected flux peaks in the optical waveband but decrease significantly at longer wavelengths. However, Rayleigh scattering in the planetary atmosphere makes the reflected spectra comparatively steeper. The geometric albedo also decreases with the increase in wavelength because of the same reason i.e. Rayleigh scattering. The amount of reflected flux for the planets orbiting {M dwarfs} is significantly less compared to {the stars of} F, G and K spectral types. This is because the input stellar spectra peaks in the infrared wavelength region where Rayleigh scattering is negligible.
The absorption lines of the {biosignatures} like O$_2$, H$_2$O, O$_3$, etc. are dominant in the geometric albedo. Owing to the fact that prebiotic early {Earth-like exoplanets} have a greater percentage of greenhouse gas CO$_2$, {they scatter} more radiation than the present {Earth-like exoplanets do}. A present Earth-like exoplanet with higher abundance of greenhouse gases also have greater albedo.
We have also estimated the maximum possible values of Bond albedo for the known exoplanets and thus given a limit on Bond albedo for the planets to remain habitable.

We also investigated the effects of surface Bond albedo on the reflected spectra and geometric albedo for various solid and liquid surface composition. We considered several kinds of solid and liquid surfaces e.g., (1) present {Earth-like} surface composition, (2) early {Earth-like} surface composition, (3) 100$\%$ ocean cover, (4) 50$\%$ ocean and remaining with trees and grass and (5) 83$\%$ ocean and remaining with snow. The reflected flux and the geometric albedo increases with the increase in surface albedo. It is minimum for no surface albedo at all. The effect of the surface albedo becomes negligible for an atmosphere thick enough to obstruct the incident stellar radiation to reach the solid or liquid surface. Thus, surface composition plays a key role in determining the reflectivity of the planet.
In the {infrared} region, the planetary surface with ocean, vegetation, desert etc. play important role in determining the reflected as well as the re-emitted thermal radiation. However we {did not} considered the re-emitted thermal radiation here.

In the second part of the work, we {presented} the transmission spectra for present {and prebiotic} Earth-like exoplanets. Since the transmission depth increases at the shorter wavelength due to scattering, an increase in the {abundance} of greenhouse {gases} yields into greater transmission depth. Also, the transmission depth reduces in magnitude with the increase in the height of the cloud level. Since the {\tt Exo-Transmit} code does not incorporate diffused radiation by scattering, it just reduces the transmission depth. As the height of the cloud increases, the threshold pressure decreases. As a consequence, a comparatively smaller atmospheric region above the clouds yields a featureless transmission spectra. Since we assume a vertically homogeneous {atmospheric} abundance, the spectral feature remain the same{, but the magnitude will change,} with the change in the cloud height.

In the future, since many big-budget missions are coming like {Habitable Worlds Observatory (HWO)}, GMT, {Thirty Meter Telescope (TMT)}, {Extremely Large Telescope (ELT)}, etc., our models will play an important role in the habitability study of the Earth-like exoplanets.
{By knowing their reflectivity, Bond albedo and the transmission spectra, we would be able to know about the 
factors like planet's surface and atmospheric composition, atmospheric $T$-$P$ profile, presence of clouds, greenhouse gases, etc., which play a key role in determining the habitable planet.}

\section*{Acknowledgements}
We acknowledge the referee José A. Caballero for giving insightful comments and the improvement in the presentation of the paper.
We thank Sukrit Ranjan for kindly providing model spectra for prebiotic Earth orbiting solar type of star and for many useful discussions. We also thank Adam Burgasser for providing the observed {near infrared} spectrum of {TRAPPIST}-1. MS would like to acknowledge Soumya Sengupta for fruitful discussions.

\newpage
\bibliographystyle{elsarticle-harv} 
\bibliography{main}
 
\label{lastpage}

\end{document}